\documentclass[10pt,journal,compsoc]{IEEEtran}
\usepackage{epsfig,endnotes}
\usepackage{comment}
\usepackage{algorithmicx, algpseudocode}
\usepackage{algorithm, amsmath}
\usepackage{enumitem}
\usepackage{amsfonts,multirow}
\usepackage{subcaption,adjustbox,tikz}
\usepackage{url, tabularx}

\setlist{nolistsep}

\makeatletter
\algrenewcommand\ALG@beginalgorithmic{\small}
\newcommand{\algmargin}{\the\ALG@thistlm}
\makeatother
\algrenewcommand{\algorithmiccomment}[1]{\footnotesize $//$#1 \small}
\algnewcommand{\parState}[1]{\State%
	\parbox[t]{\dimexpr\linewidth-\algmargin}{\strut #1\strut}}

\newcommand*\numbercircled[1]{\tikz[baseline=(char.base)]{
		\node[shape=circle,draw,inner sep=0.5pt] (char) {#1};}}

\ifCLASSOPTIONcompsoc
  \usepackage[nocompress]{cite}
\else
  \usepackage{cite}
\fi
\ifCLASSINFOpdf
\else
\fi
\begin{document}
%
\title{Circ-Tree: A B+-Tree Variant with Circular Design for Persistent Memory}
%
%
%
%

\author{Chundong~Wang,
        Gunavaran~Brihadiswarn,
        Xingbin~Jiang,
        and~Sudipta~Chattopadhyay
\IEEEcompsocitemizethanks{\IEEEcompsocthanksitem C. Wang, X. Jiang, and S. Chattopadhyay
	are with Singapore Universtiy of Technology and Design, Singapore
	\protect\\
E-mail: cd\_wang@outlook.com, xingbin\_jiang@sutd.edu.sg, and sudipta\_chattopadhyay@sutd.edu.sg.
\IEEEcompsocthanksitem G. Brihadiswarn is with University of Moratuwa,
Sri Lanka. This work was done when he worked as an intern in Singapore University of Technology and Design, Singapore.
	\protect \\
Email: gunavaran.15@cse.mrt.ac.lk.
}
\thanks{Manuscript received MM DD, 2019; revised MM DD, 2019.}}

%
%

\markboth{IEEE Tranasctions on XXXXXX}%
{Circ-Tree: A B+-Tree Variant with Circular Design for Persistent Memory}
%



\IEEEtitleabstractindextext{%
\begin{abstract}
Several B+-tree variants have been developed to exploit
the performance potential of byte-addressable non-volatile
memory (NVM). In this paper,
we attentively investigate the properties
of B+-tree and find that,
a conventional B+-tree node is a linear structure in which
key-value (KV) pairs are maintained from the zero offset
of the node. These pairs are shifted in a unidirectional
fashion for insertions and deletions.
Inserting and deleting one KV pair may inflict a large amount
of write amplifications due to shifting KV pairs. This badly
impairs the performance of in-NVM B+-tree. In this paper, we
propose a novel {\em circular} design for B+-tree. With
regard to NVM's byte-addressability, our Circ-tree design
embraces tree nodes in a circular structure without a fixed
base address, and 
bidirectionally shifts KV pairs in a node for insertions
and deletions to minimize write amplifications.
We have implemented a prototype for Circ-Tree and conducted
extensive experiments. Experimental results show that Circ-Tree
significantly outperforms two state-of-the-art in-NVM
B+-tree variants, i.e., NV-tree and FAST+FAIR,
by up to 1.6$\times$ and 8.6$\times$, respectively, in terms of
write performance. 
The end-to-end comparison by running YCSB to KV store systems
built on NV-tree, FAST+FAIR, and Circ-Tree
reveals that Circ-Tree yields up to 29.3\% and 47.4\% higher write performance, respectively,
than NV-tree and FAST+FAIR.
\end{abstract}

}

\maketitle

\IEEEdisplaynontitleabstractindextext

%
\IEEEpeerreviewmaketitle

\IEEEraisesectionheading{\section{Introduction}\label{sec:introduction}}

The next-generation non-volatile memory (NVM) has DRAM-like 
byte-addressability and disk-like durability. Computer 
architects have proposed to place NVM on the memory bus 
alongside DRAM to build {\em persistent memory}~\cite{NV:Klin:2013,NVM:B+-tree:TRIOS-2013,FS:NOVA:FAST-2016,FS:PMFS:EuroSys-2014}.
However, the write latency of NVM technologies is generally 
longer than that of DRAM~\cite{FS:HiNFS:Eurosys-2016, NVM:HiKV:ATC-2017, kv:NoveLSM:ATC-2018, NVM:FAST+FAIR:FAST-2018}. 
Moreover, to guarantee crash consistency of data  maintained in NVM
demands the execution of cache line flush and memory fence, as the 
CPU or the memory controller may 
alter the 
writing order of multiple writes from CPU cache to memory~\cite{NVM:NV-tree:FAST-2015, NVM:FAST+FAIR:FAST-2018, NVM:loose-order:ICCD2014, NVM:Duet:ASPLOS-2014, NVM:log-nvmm:ATC-2017, NVM:soft-updates:ATC-2017}. 
Despite being effective in preserving a desired writing order, 
the usage of cache line flushes and memory fences incurs further 
performance overhead for in-NVM data structures.

B+-tree, as one of the critical building blocks for computer systems,
has been tailored to persistently store and manage key-value (KV) 
pairs in NVM~\cite{NVM:CDDS-tree:2011, NVM:NV-tree:FAST-2015, NVM:wB+-tree:VLDB-2015, NVM:FPTree:SIGMOD-2016, NVM:FAST+FAIR:FAST-2018,btree:Crab:LCTES-2019}. 
These in-NVM B+-tree variants attempt to guarantee crash consistency 
with minimal performance overheads
through either 
managing unsorted KV pairs of a node in an append-only fashion~\cite{NVM:NV-tree:FAST-2015, NVM:wB+-tree:VLDB-2015, NVM:FPTree:SIGMOD-2016}, or leveraging architectural supports to orderly 
shift sorted KV pairs with the reduction of cache line flushes~\cite{NVM:FAST+FAIR:FAST-2018}. 
Despite employing different techniques, 
all these works inherit 
an important property from standard B+-tree: 
their tree nodes are organized in a {\em linear} structure that starts at the zero offset and 
spans over a contiguous space.
In particular, for a B+-tree with sorted KV pairs, inserting and deleting 
a KV pair always shift keys in a {\em unidirectional} fashion, i.e., to 
the right for insertion and to the left for deletion, given keys in ascending 
order. Assuming that the KV pair under insertion is with
the smallest key of a node, 
all existing KV pairs in the node have to be shifted. 
Consequently, a large amount of memory writes with 
cache line flushes and memory fences are to take place.

B+-tree has been used since the era of hard disk.
The linear tree node is perfectly favored by a 
hard disk which rotates a magnetic head to sequentially 
write and read data. 
However, the byte-addressability entitles NVM the very access 
flexibility on manipulating data structures stored in it. 
To this end, we propose a novel {\em circular} tree node
and design a {\bf Circ-Tree} for NVM based on our proposed 
node structure. The main ideas of Circ-Tree are summarized 
as follows.
\begin{itemize}
	\item Circ-Tree organizes sorted KV pairs in a circular 
	node structure that has no fixed base address. Upon an 
	insertion or deletion, Circ-Tree shifts existing KV pairs 
	of a node in a  
	{\em bidirectional} manner.
	\item Circ-Tree decides whether to shift to the left or 
	to the right for an insertion or deletion by considering 
	which direction would generate fewer shifts of existing 
	KV pairs. 
	This is to reduce memory writes to NVM via cache line 
	flushes and memory fences.
\end{itemize}

With the novel circular design, Circ-Tree
reduces write amplifications caused by shifting KV pairs while retaining
good read performance with respect to sorted KV pairs.
We have designed and implemented a prototype of Circ-Tree 
and performed extensive experiments.
Evaluation results 
show that the write performance of Circ-Tree
can be up to 1.6$\times$ and 8.6$\times$ that of NV-tree and FAST+FAIR,
respectively. 

We have also built KV store systems with three B+-tree variants 
(i.e., Circ-Tree, NV-Tree, and FAST+FAIR) to 
observe their end-to-end performances. 
With YCSB~\cite{YCSB}, Circ-tree yields 29.3\% and
47.4\% higher write performance than NV-tree and 
FAST+FAIR, respectively, with all of them directly 
committing large values into NVM.

The remainder of this paper is organized as follows. 
In Section~\ref{sec:bg}, we brief the background of NVM and 
state-of-the-art B+-tree variants for NVM.
We show a motivating example in Section~\ref{sec:mot}.
We detail the design of Circ-Tree in Section~\ref{sec:circ} and Section~\ref{sec:op}.
We present evaluation results in Section~\ref{sec:evaluation}
and conclude this paper in Section~\ref{sec:conclusion}.

\section{Background and Related Works}\label{sec:bg}

NVM technologies, such as spin-transfer torque RAM (STT-RAM), phase change memory (PCM), and
3D XPoint, have become prominent with DRAM-like byte-addressability and disk-like durability. 
Compared to DRAM, NVM generally has asymmetrical write/read speeds, 
especially with longer write latencies~\cite{NVM:NV-tree:FAST-2015,NVM:Quartz:Middleware-2015,FS:HiNFS:Eurosys-2016,NVM:FAST+FAIR:FAST-2018, kv:NoveLSM:ATC-2018,kv:hash:OSDI-2018}.

One way to incorporate NVM in a computer system 
is to build persistent memory by putting NVM alongside DRAM on the memory bus.
Though,
new challenges emerge when a data structure is ported from hard disk
to persistent memory that is directly operated by a CPU.
First, modern CPUs mainly support an atomic write of 8B 
~\cite{NVM:Intel:16B,FS:PMFS:EuroSys-2014,NVM:Tinca:SC-2017,NVM:FAST+FAIR:FAST-2018,FS:backpointer:FAST-2012}.
In addition, the exchange unit between CPU cache 
and memory is a cache line that typically has a size of 64B or 128B.  
Secondly, for multiple store operations from CPU cache to memory,
the CPU or the memory controller may perform them in an order 
differently from the programmed order. 
Such reordered writes are 
detrimental to the crash consistency of in-NVM data structures.  
For example, the allocation of a new object must be completed before
the pointer to the object is recorded.
If the writing order is reversed but a crash 
occurs, the pointer might turn to be dangling.
Using cache line flushes  
and memory fences, e.g., {\tt clflush} and {\tt mfence} in the x86 architecture,
is an effective method to 
retain a desired writing order. Cache line flush 
explicitly flushes a cache line
to memory. Memory fence makes a barrier to regulate that
memory operations after the barrier cannot proceed unless
ones before the barrier complete.
However, the execution of 
cache line flushes and memory fences
incurs performance 
overheads~\cite{NVM:NV-tree:FAST-2015, NVM:FAST+FAIR:FAST-2018, NVM:loose-order:ICCD2014, NVM:Duet:ASPLOS-2014, NVM:log-nvmm:ATC-2017}. Recently,
Intel has introduced new instructions, i.e., {\tt clflushopt} and {\tt clwb},
to replace {\tt clflush} with reduced performance overheads.

Computer scientists have proposed a number of artifacts for 
system- and application-level softwares to utilize persistent memory 
on the memory bus~\cite{FS:BPFS:SOSP-2009, NVM:NV-Heaps:ASPLOS-2011, NVM:Mnemosyne:ASPLOS2011, NVM:B+-tree:TRIOS-2013, fs:unioning:2013, FS:PMFS:EuroSys-2014, NVM:Mojim:APLOS2015, NVM:ThyNVM:Micro-2015, NVM:shredder:ASPLOS-2016, FS:NOVA:FAST-2016, NVM:fine-grained-metadata:MSST-2016, NVM:HiKV:ATC-2017, NVM:log-nvmm:ATC-2017, NVM:Tinca:SC-2017, kv:NoveLSM:ATC-2018, kv:hash:OSDI-2018,NVM-log-free-ATC2018}.
In particular, several in-NVM B+-tree variants have been developed~\cite{NVM:CDDS-tree:2011,NVM:unsorted-node-PCM:CIDR-2011,NVM:NV-tree:FAST-2015,NVM:wB+-tree:VLDB-2015,NVM:FPTree:SIGMOD-2016,NVM:FAST+FAIR:FAST-2018,btree:Crab:LCTES-2019}.
CDDS-Tree keeps all nodes sorted in NVM~\cite{NVM:CDDS-tree:2011}. 
It calls cache line flushes and memory fences 
while shifting every KV pairs to orderly write modified data back to NVM. 
Yang et al.~\cite{NVM:NV-tree:FAST-2015} found that, 
for CDDS-Tree, the cost of using {\tt clflush} and {\tt mfence} 
to sort its leaf nodes might take up to 90\% of overall consistency cost.
Accordingly they studied the idea of unsorted tree
nodes~\cite{NVM:unsorted-node-PCM:CIDR-2011} and
proposed NV-tree. NV-tree only enforces crash consistency to leaf nodes
and manages them in an unsorted fashion. 
It appends a KV pair to a leaf node with respective labels for insertion,
update and deletion. 
However, for every write or read request,
NV-tree has to scan all unsorted 
KV pairs in a leaf node to determine the existence of the key
under insertion/deletion/search.
Oukid et al.~\cite{NVM:FPTree:SIGMOD-2016} looked into this issue and proposed 
FPTree. A leaf node of FPTree includes a hash value for stored keys,
which can be first checked in order to accelerate searching over unsorted KV pairs.

Hwang et al.~\cite{NVM:FAST+FAIR:FAST-2018} reconsidered sorted 
B+-tree nodes and proposed FAST+FAIR that 
exploits the store dependencies in shifting a sequence of KV pairs 
in a node on insertion/deletion (e.g., $KV_i\xrightarrow[]{\text{store}}KV_{i+1}, KV_{i-1}\xrightarrow[]{\text{store}}KV_{i}$, etc.).
Store dependencies  
 impose a natural writing order among KV pairs
and result in reducing the cost of 
using cache line flushes and memory
fences to forcefully write back modified KV pairs to NVM~\cite{arch:persist-barriers:Micro-2015, NVM:ordering:Micro-2016, arch:TSO:2017}.
Nevertheless, FAST+FAIR still needs to orderly flush dirty cache lines 
holding shifted KV pairs to prevent them from being written to NVM in 
an altered order.

\section{A Motivational Example}\label{sec:mot}

State-of-the-art B+-tree variants developed for NVM inherit an important property 
from the standard B+tree:
a B+-tree node is a linear structure that starts at a fixed base address, i.e., zero offset, and
spans over a contiguous space; as a result, insertions and deletions onto a B+-tree node
always shift KV pairs in a unidirectional fashion. 
However, in this paper, we show that the linear structure of B+-tree 
can be logically viewed in a circular fashion with novel 
operational strategies. We will first introduce a simple example to 
illustrate the key motivation behind such a design. 

\begin{figure}[t]
	\centering
	\begin{subfigure}[t]{0.75\columnwidth}
		\includegraphics[width=\textwidth]{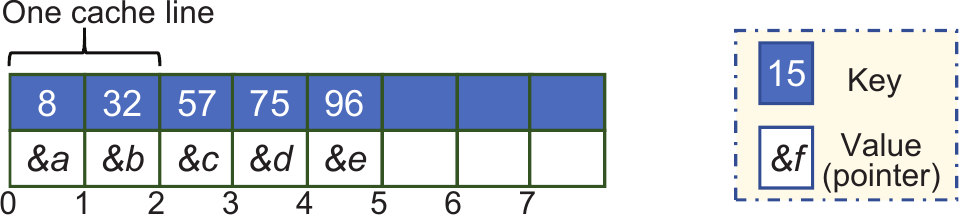}
		\caption{A B+-tree node in the linear structure}
		\label{fig:linear-node}
	\end{subfigure}\vspace{0.5ex}	\\
	\begin{subfigure}[t]{0.48\columnwidth}
		\adjincludegraphics[valign=t, width=\textwidth]{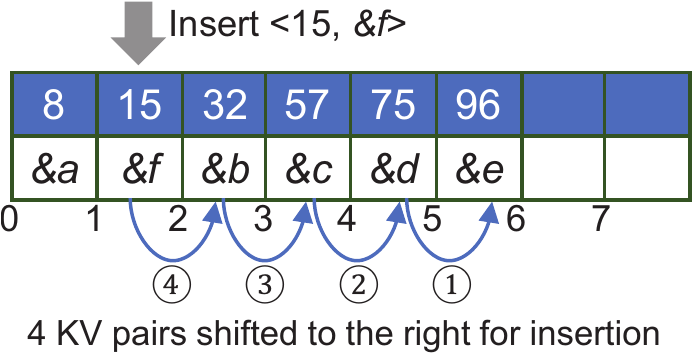}	
		\caption{Insertion to a linear node}\label{fig:old-insert}
	\end{subfigure}	
	\hfill
	\begin{subfigure}[t]{0.48\columnwidth}
		\adjincludegraphics[valign=t, width=\textwidth]{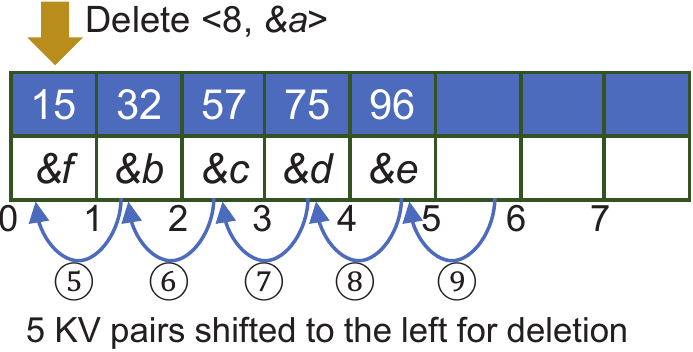}			
		\caption{Deletion from a linear node}\label{fig:old-delete}
	\end{subfigure}	\\ \vspace{1.5ex}
	\begin{subfigure}[t]{0.48\columnwidth}
		\adjincludegraphics[valign=t, , width=\textwidth]{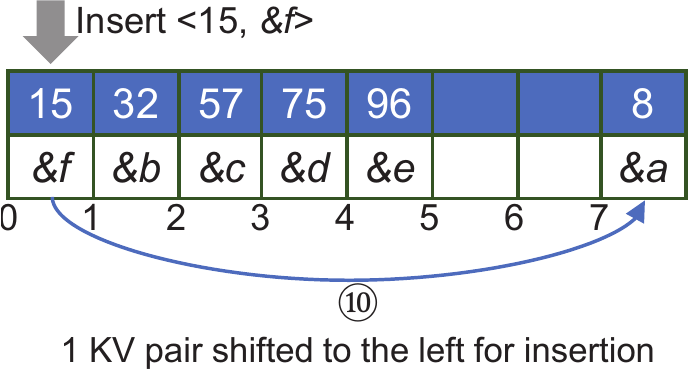}				
		\caption{Insertion to a circular node}\label{fig:new-insert}
	\end{subfigure}		
	\hfill
	\begin{subfigure}[t]{0.48\columnwidth}
		\adjincludegraphics[valign=t, , width=\textwidth]{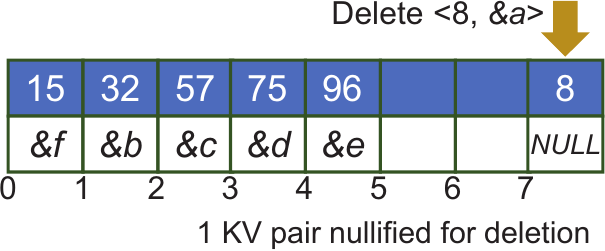}						
		\caption{Deletion from a circular node}\label{fig:new-delete}
	\end{subfigure}
	\vspace{-1ex}		
	\caption{An Example to Compare Classic Linear B+-tree Node to Circ-Tree's Circular Node Design}~\label{fig:treenode}
	\vspace{-3ex}
\end{figure}

Figure~\ref{fig:linear-node} shows a linear B+-tree node that keeps
keys in ascending order.
Without loss of generality, 
we assume that one CPU cache line holds two KV pairs. 
In Figure~\ref{fig:old-insert}, we insert a 
KV pair $\langle15, \&f\rangle$ into the node.
Since its key would be the second smallest of the node, four KV pairs must 
be shifted to the right in order to keep the node sorted.  
Considering store dependencies in shifting KV pairs,
we need to call three cache line flushes and memory fences. This is because three cache lines 
from the zero offset are modified due to shifting KV pairs and 
inserting the new KV pair.
In other words, inserting $\langle15, \&f\rangle$ causes 
writing other four KV pairs with flushing three cache lines. 
When we delete the smallest key of the node as shown in Figure~\ref{fig:old-delete},
 all greater KV pairs must be shifted to the left.
Again, three dirty cache lines are involved in the deletion and they
must be orderly flushed.

The linear structure is well fit for hard disk with a magnetic head rotating to write/read data. 
The byte-addressable NVM, as directly operated by 
the CPU, provides an opportunity to 
revolutionize the linear node structure for B+-tree.
Let us assume that in Figure~\ref{fig:old-insert} we do 
not shift four greater KV pairs to the right.
Instead we shift the smallest $\langle8, \&a\rangle$ to the rightmost end of the node and insert $\langle15, \&f\rangle$
in the leftmost slot (cf. Figure~\ref{fig:new-insert}). By shifting one KV pair, 
just the leftmost and the rightmost cache lines are modified and flushed.
As to the deletion in Figure~\ref{fig:old-delete}, now that $\langle8, \&a\rangle$ 
has been shifted to the rightmost cache line of the node, we nullify its value 
to be $\mathsf{NULL}$ without shifting any KV pair (cf. Figure~\ref{fig:new-delete})
and hence only one cache line needs to be flushed.

As observed from the preceding example, we transform the linear node of B+-tree
into a {\em circular} structure that supports 
{\em bidirectional} shifting. 
Such a circular design significantly reduces write amplifications caused by 
shifting KV pairs while retaining sorted B+-tree nodes to support fast lookup.
This is the main motivation driving the design of our Circ-Tree.

\section{Tree Node with Circular Deign}\label{sec:circ}
\subsection{Circular B+-tree Node}\label{sec:node}

\begin{figure}[t]		
	\centering
	\scalebox{1.00}{\includegraphics[width=\columnwidth]{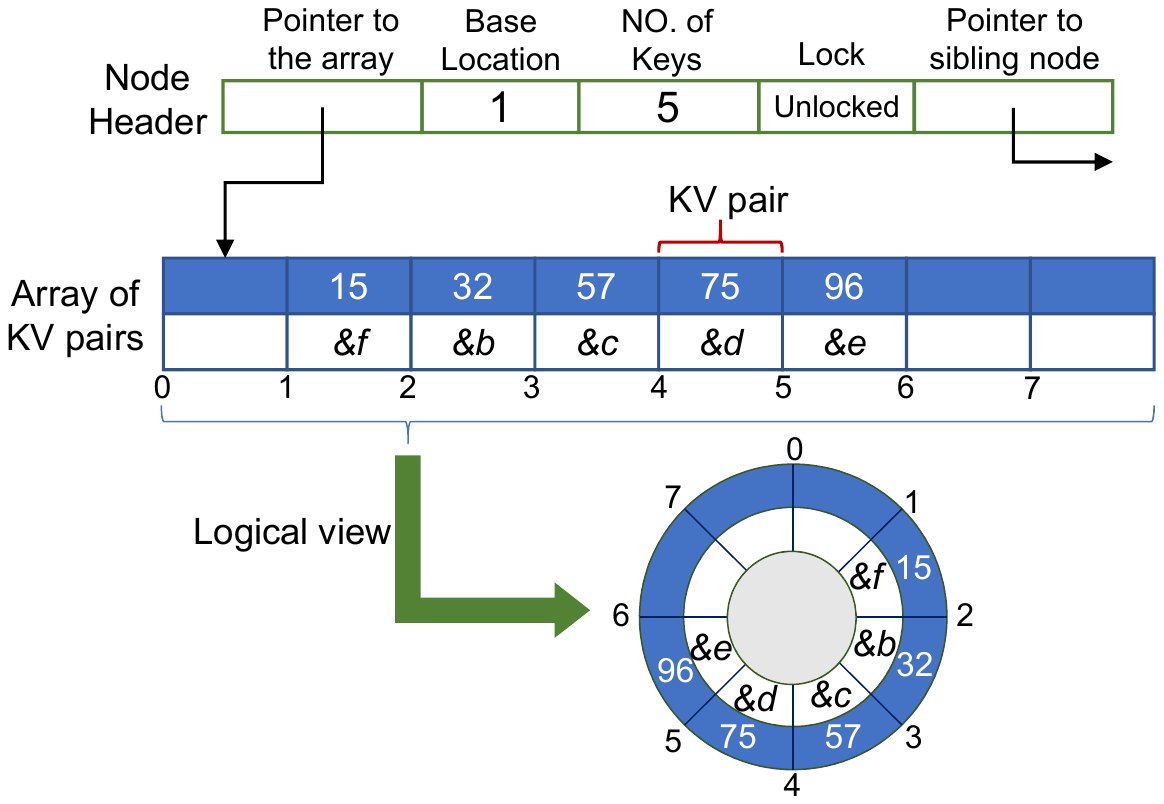}} 
	\vspace{-3ex}
	\caption{The Circular Structure of Circ-Tree's Leaf Node}~\label{fig:node}
	\vspace{-3ex}	
\end{figure}

Circ-Tree views a linear node as a circular buffer~\cite{circular-buffer}
that no longer has a fixed base address at the zero offset.
Figure~\ref{fig:node} instantiates one leaf node (LN) of Circ-Tree.
An LN is composed of two parts, i.e., an array of KV pairs and a node header.
The reason why Circ-Tree separates its tree node into a header and an array of KV pairs
is explained in Section~\ref{sec:opt}.
As to the node header, it contains the following items:
\begin{enumerate}[label=\arabic*)]
	\item a pointer (8B) pointing to the array of KV pairs,
	\item the current base location (4B),
	\item the number of valid keys in the LN (4B),
	\item a lock (8B) for concurrent access, and
	\item a pointer (8B) pointing to the right sibling LN. 
\end{enumerate}
The base location indicates where the smallest key is stored. The number of keys
maintains the boundary of circular space that is being filled with valid KV pairs.
We can use an 8B atomic write to atomically modify these two items together as well as either pointer in
the node header.
Figure~\ref{fig:node} also illustrates a logically circular view of the example LN.

An internal node (IN) of Circ-Tree has the same structure as its LN,
except that each IN has an unused key so that the number of values is
always one more than the number of keys. This is in line with the 
definition of B+-tree.
In the LN, the value of a KV pair is a pointer to a record where the actual value
is stored. 
In the IN, the value is a pointer to the lower-level IN or LN.
Circ-Tree enforces crash consistency to all INs and LNs
and it incorporates strict modification orders in insertion and 
deletion for crash recoverablity.

\subsection{Optimization for Circular Design}\label{sec:opt}

The circular node structure enables Circ-Tree to shift KV pairs bidirectionally, which
can be implemented using 
the modulo operation ($\%$ is the modulo operator in C/C++/Python/Java).
Let the maximum number of values that can be stored in a node
be $\mathsf{N}$. Assuming that a new KV pair should be inserted at the $i$-th offset from
the base location ($b$), the real position $p$ in the array of KV pairs would be
\begin{equation*}
p = (b + i) \% \mathsf{N}.
\end{equation*}
The left and right moves by one for the base location would be $(b-1)\%\mathsf{N}$ and $(b+1)\%\mathsf{N}$, respectively.

In practice, the modulo operation is expensive for a CPU because it is a 
form of integer division. As Circ-Tree frequently and bidirectionally shifts KV pairs
for insertions and deletions, we need an efficient calculation to substitute modulo
operations. We find that, when $\mathsf{N}$ is a power of two, i.e., $\mathsf{N} = 2^{m}$ ($m > 0$), the modulo operation
can be reformed as 
\begin{equation*}
p = (b + i) \% \mathsf{N} = (b + i) \& (\mathsf{N} - 1),
\end{equation*}
in which $\&$ is the bitwise AND operator. For example, given $\mathsf{N} = 256$, $(133 + 165) \% 256 = (133 + 165) \& 255 = 42$.
The bitwise AND operation is much more CPU-friendly than the modulo operation.
As a result, we organize 
an array that has a size in a power of two, e.g., $2^6$, for KV pairs
and is competent for using bitwise AND operation for modulo calculations.
As a CPU cache line usually has a size of 64B or 128B,
we separate the array of KV pairs from the node header to make the former cache 
line-aligned and-friendly.
Meanwhile, the node header has an overall size of 32B and a multiple of it can be
fitted in one cache line. This further improves the CPU cache efficiency of Circ-Tree.

\section{Search, Insertion, \& Deletion}\label{sec:op}

\begin{figure}[t]		
	\centering
	\begin{subfigure}[t]{0.475\columnwidth}
		\includegraphics[width=\textwidth]{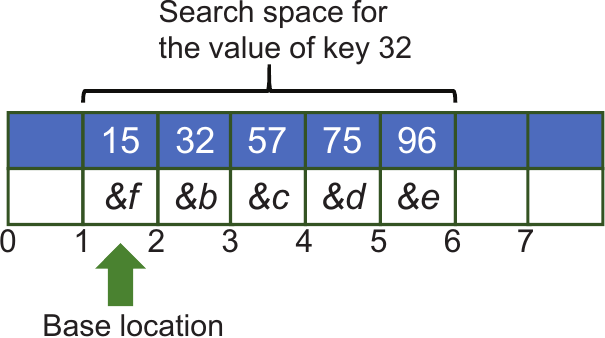}
		\caption{Searching in the contiguous space (Case 1)}\label{fig:search-a}
	\end{subfigure}
	\hfill
	\begin{subfigure}[t]{0.475\columnwidth}
		\includegraphics[width=\textwidth]{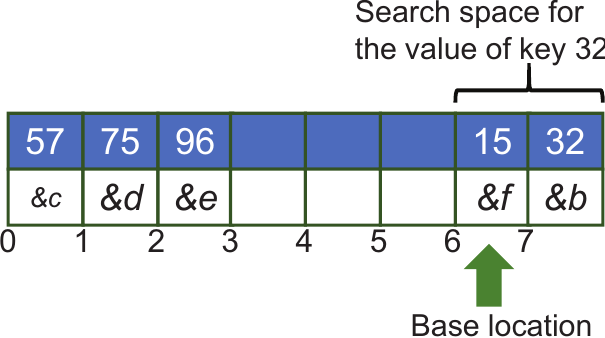}		
		\caption{Searching in the disjointed space (Case 2)}\label{fig:search-b}
	\end{subfigure}
	\vspace{-1ex}		
	\caption{An Example of Searching for a Key by Circ-Tree}~\label{fig:search}
		\vspace{-2ex}		
\end{figure}

\subsection{Search}\label{sec:search}

Search is critical for B+-tree because it is essential
part of insertion and deletion.
Since Circ-Tree employs a logically circular node structure, 
valid KV pairs of an LN can occupy either one contiguous space or two segments.
Figure~\ref{fig:search} exemplifies both possible cases.
To avoid cache misses due to traversing disjointed cache lines as shown in Figure~\ref{fig:search-b},
Circ-Tree only searches in a contiguous space. 
First, whether the distance between locations of the smallest and greatest keys
is positive or negative indicates which case in Figure~\ref{fig:search} 
a node corresponds to.
Secondly, with two disjointed segments,
given a key to be searched, e.g., 32 in Figure~\ref{fig:search-b},
Circ-Tree decides which segment to be searched by
comparing the key under search to the key at the zero offset.
Therefore, as shown in Figure~\ref{fig:search-a} and Figure~\ref{fig:search-b}, 
Circ-Tree manages to search only one continuous space in both cases
without jumping between discontinuous cache lines.
Circ-Tree can use either binary or linear search. Because linear search
is more CPU cache-friendly than binary search especially with short arrays~\cite{NVM:FAST+FAIR:FAST-2018}, 
Circ-Tree employs the former in its implementation.

\subsection{Insertion}\label{sec:insert}

Algorithm~\ref{algo:insert} captures
the single-threading procedure of insertion for Circ-Tree with the x86 architecture.
Inserting a new KV pair 
starts by traversing from the tree root until reaching the target LN (Line~\ref{algo:line:search}). 
Then Circ-Tree 
checks if the current LN is full (Line~\ref{algo:line:to-split}). 
A full LN is split 
with the newly-arrived KV pair (Line~\ref{algo:line:split}). 
Otherwise, Circ-Tree inserts the new KV pair into the LN (Lines~\ref{algo:line:insert-start}
to~\ref{algo:line:insert-end}).

\begin{algorithm}[htbp]	
	\caption{Insertion of Circ-Tree ({\tt Insert}($<k, v>$))}\label{algo:insert}
	\begin{algorithmic}[1]
		\Require A KV pair $<k, v>$ to be inserted 
		\State Search from the root until the target node header $nh$ and KV array $A$\Comment{$nh$ has the base location $b$ and the number of KV pairs $n$} \label{algo:line:search}
		\If {($nh.n \ge \mathsf{N}$)} \label{algo:line:to-split}
			\State {\tt split}($nh$,\ \ $<k, v>$); \Comment{To split the node with $<k, v>$}\label{algo:line:split}
		\Else \label{algo:line:insert-start}
			\If {($k < A[(nh.b + \frac{nh.n}{2}) \& (\mathsf{N} - 1)].key$)} \label{algo:line:shift-direction}
				\For{($i := 0; i <\frac{nh.n}{2};$ $i$ := $i+1$)}
					\State $index := (nh.b + i)\ \&\ (\mathsf{N} - 1)$; \label{algo:line:move-start}
					\If {($k > A[index].key$)} \label{algo:line:smaller-than-newkey}
						\State $A[(index - 1)\ \&\ (\mathsf{N - 1})].val$ := $A[index].val$;
						\State $A[(index - 1)\ \&\ (\mathsf{N} - 1)].key$ := $A[index].key$;
						\If {($A[index]$ is at the start of a cache line)}\label{algo:line:flush-cacheline1-start}
							\State {\tt Flush\_cacheline}($\&A[index - 1]$);
						\EndIf \label{algo:line:move-end}\label{algo:line:flush-cacheline1-end}
					\Else
						\State {\bf break}; \Comment{Find the appropriate position} \label{algo:line:position1}
					\EndIf
				\EndFor
				\State $A[(nh.b + i - 1)\ \&\ (\mathsf{N - 1})].key$ := $k$;  \label{algo:line:insert-place-start}
				\State $A[(nh.b + i - 1)\ \&\ (\mathsf{N - 1})].val$ := $v$;
				\State {\tt Flush\_KV}($\&A[(nh.b + i - 1) \& \mathsf{N - 1}]$); \label{algo:line:insert-place-end}		
				\State {\tt Update\_b\_n}($nh$, $((nh.b - 1) \ \&\ (\mathsf{N - 1}))$, $(nh.n + 1)$);\label{algo:line:update-bn1-start}
				\State {\tt Flush\_b\_n}($\&nh$);\label{algo:line:update-bn1-end}
			\Else
				\For {($i := nh.n - 1; i>= \frac{nh.n}{2};$ $i$ := $i-1$)} \label{algo:line:right-start}
					\State $index := nh.b + i\ \&\ (\mathsf{N} - 1)$;\label{algo:line:right-move-start}
					\If {($k < A[index].key$)}
						\State $A[(index + 1)\ \&\ (\mathsf{N - 1})].val$ := $A[index].val$;
						\State $A[(index + 1)\ \&\ (\mathsf{N - 1})].key$ := $A[index].key$;
						\If {($A[(index + 1)\&(\mathsf{N - 1})]$ is at the start of \indent\indent\indent\indent\indent\indent a cache line)}
							\State \parbox[t]{\dimexpr\linewidth-\algorithmicindent}{{\tt Flush\_cacheline}($\&A[(index+1)\&(\mathsf{N-1})]$);\strut}
						\EndIf\label{algo:line:right-move-end}
					\Else
						\State {\bf break}; \Comment{Find the appropriate position}					
					\EndIf
				\EndFor
				\State $A[(nh.b + i + 1)\ \&\ (\mathsf{N - 1})].key$ := $k$; \label{algo:line:right-put-start}		
				\State $A[(nh.b + i + 1)\ \&\ (\mathsf{N - 1})].val$ := $v$;
				\State {\tt Flush\_KV}($\&A[(nh.b + i + 1) \& \mathsf{N - 1}]$);\label{algo:line:right-put-end}	
				\State {\tt Update\_b\_n}($nh$, $\mathbf{NIL}$, $(nh.n + 1)$); \label{algo:line:n-right-increase}
				\State {\tt Flush\_b\_n}($nh$);	\label{algo:line:right-end}			
			\EndIf	
		\EndIf \label{algo:line:insert-end}
	\end{algorithmic}
\end{algorithm}

Next, Circ-Tree needs to decide the direction to shift KV pairs
(Lines~\ref{algo:line:shift-direction}).
Circ-Tree compares the new key to the key at the middle position. If the new key is
smaller than the middle one, then the new key is inserted into the logically smaller
half. Shifting to the left incurs fewer KV pairs to be moved, because shifting to the right
surely moves more than half KV pairs. 
Circ-Tree starts shifting KV pairs which have smaller keys than the new key (Line~\ref{algo:line:smaller-than-newkey})
to the left (Lines~\ref{algo:line:move-start} to~\ref{algo:line:move-end}) until it reaches the first key
that is not smaller than the new key (Line~\ref{algo:line:smaller-than-newkey}). This is the position where the newly-arrived KV pair is placed (Line~\ref{algo:line:position1}).
When shifting KV pairs to the left,  if 
Circ-Tree crosses a boundary of cache lines and enters 
the next left cache line, it flushes the modified dirty cache line for consistency and persistency (Lines~\ref{algo:line:flush-cacheline1-start} to~\ref{algo:line:flush-cacheline1-end}).
After inserting and flushing the new KV pair into the appropriate position (Lines~\ref{algo:line:insert-place-start} to~\ref{algo:line:insert-place-end}),
the base location of the LN is moved to the left by one (i.e., $((nh.b - 1) \ \&\ (\mathsf{N - 1}))$)
and the number of KV pairs is increased by one (i.e., $(nh.n + 1)$). 
As Circ-Tree keeps these two items in
an 8B word, it 
atomically modifies and flushes the 8B to complete the insertion (Lines~\ref{algo:line:update-bn1-start} to~\ref{algo:line:update-bn1-end}). In other words, 
Circ-Tree leverages this 8B atomic write as the end mark of insertion, and it is useful in crash recovery.

If the new key falls into the logically greater half, Circ-Tree shifts KV pairs to the right (Lines~\ref{algo:line:right-start} to~\ref{algo:line:right-end}).
Circ-Tree moves KV pairs that embrace greater keys than the key to be inserted,
and flushes dirty cache lines where necessary (Lines~\ref{algo:line:right-move-start} to \ref{algo:line:right-move-end}). Then it inserts and flushes the newly-arrived KV pair into the proper
location (Lines~\ref{algo:line:right-put-start} to~\ref{algo:line:right-put-end}). 
These steps are similar to steps in a left shifting except that, in the end, only the number of KV pairs needs to be increased by one (Lines~\ref{algo:line:n-right-increase} to~\ref{algo:line:right-end})
as the base location does not change in a right shifting.
The increase of the number of KV pairs is also accomplished with an 8B atomic write.

\begin{figure*}[t]
	\centering
	\begin{subfigure}[t]{0.6\columnwidth}
		\includegraphics[width=\textwidth]{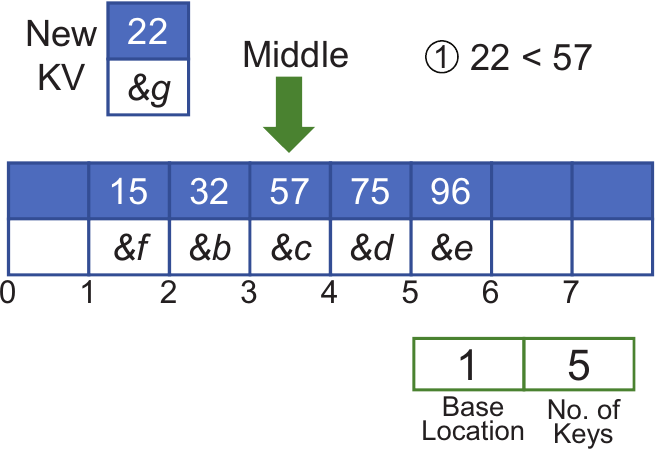}
		\caption{22 $<$ 57, $\langle 22, \&g\rangle$ to be inserted to the lower half by shifting to the left.}\label{fig:insert-a}
	\end{subfigure}
	\hfill
	\begin{subfigure}[t]{0.6\columnwidth}
		\includegraphics[width=\textwidth]{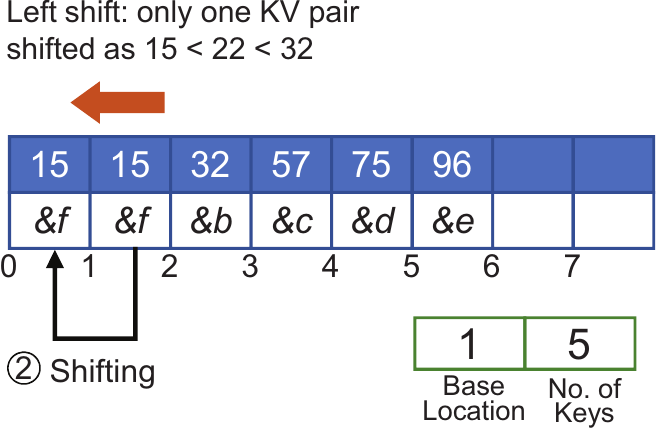}
		\caption{Shifting KV pairs that have smaller keys than the new key, i.e., only $\langle 15, \&f\rangle$.}\label{fig:insert-b}
	\end{subfigure}
	\hfill
	\begin{subfigure}[t]{0.76\columnwidth}
		\includegraphics[width=\textwidth]{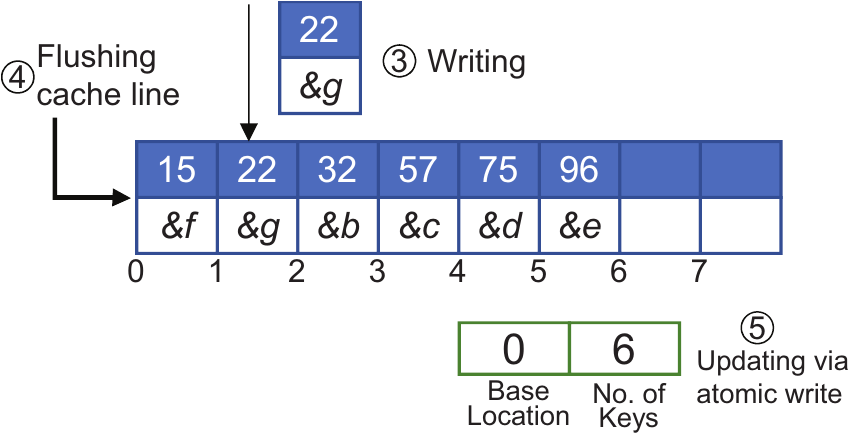}
		\caption{Writing $\langle 22, \&g\rangle$ in the proper position and flushing the corresponding cache line before atomically updating the base location and the number of keys.}\label{fig:insert-c}
	\end{subfigure}
	\vspace{-1ex}		
	\caption{An Example of Inserting a New KV Pair into a Circular LN by Circ-Tree}~\label{fig:insert}
	\vspace{-3ex}
\end{figure*}

\begin{figure*}[t]
	\centering
	\begin{subfigure}[t]{0.7\columnwidth}
		\includegraphics[width=\textwidth]{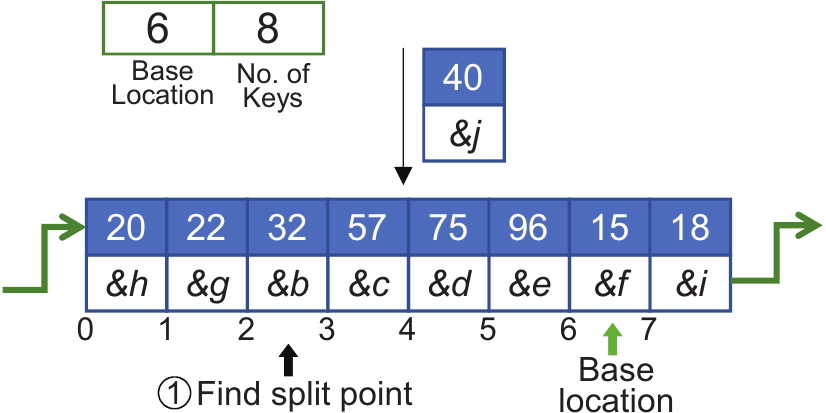}
		\caption{Finding the split point of LN and which half the new KV pair will join (\numbercircled{1})}\label{fig:split-a}
	\end{subfigure}
	\hfill
	\begin{subfigure}[t]{1.2\columnwidth}
		\includegraphics[width=\textwidth]{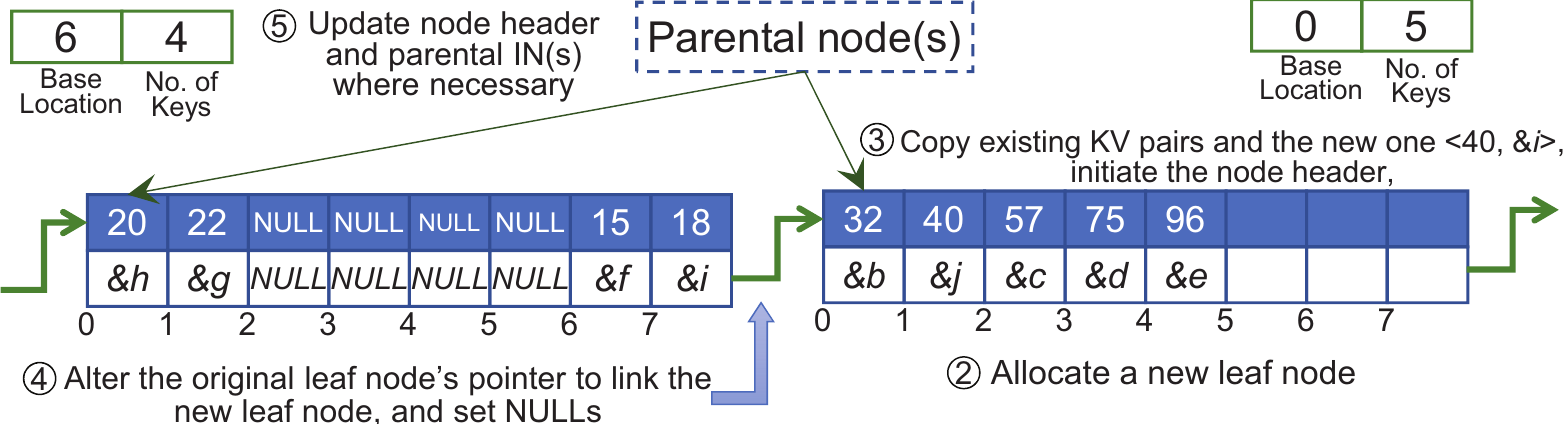}		
		\caption{Creating new LN (\numbercircled{2}), copying KV pairs (\numbercircled{3}), linking the new LN into LN linked list as well as setting $\mathsf{NULL}$s (\numbercircled{4}), and updating node header and parental INs (\numbercircled{5})}\label{fig:split-b}
	\end{subfigure}
	\vspace{-1ex}		
	\caption{An Example of Splitting an LN with a New KV pair by Circ-Tree}~\label{fig:split}
	\vspace{-3ex}
\end{figure*}

Figure~\ref{fig:insert} describes how Circ-Tree 
deals with an insertion request into the LN shown in Figure~\ref{fig:node}.
To insert the new KV pair $\langle 22, \&g\rangle$, Circ-Tree first decides the shifting direction
(\numbercircled{1} in Figure~\ref{fig:insert-a}). As the new key 22 is smaller than the key at the middle position,
shifting to the left incurs fewer moves of KV pairs. As shown in Figure~\ref{fig:insert-b}, Circ-Tree shifts one KV pair to the left to make room for $\langle 22, \&g\rangle$ (\numbercircled{2} in Figure~\ref{fig:insert-b}). Note that the shift of $\langle 15, \&f\rangle$
is inside one cache line and does not cross a cache line boundary. 
Next, Circ-Tree puts the new KV pair in the appropriate position, and flushes the modified 
cache line due to moving $\langle 15, \&f\rangle$ and writing  $\langle 22, \&g\rangle$
(\numbercircled{3} and \numbercircled{4} in Figure~\ref{fig:insert-c}). 
In the eventual step,
the base location and the number of keys is atomically modified and flushed to the NVM  (\numbercircled{5} in Figure~\ref{fig:insert-c}).

\textbf{Split}\hspace{2ex}  
Circ-Tree splits an LN when the LN's space is consumed up upon inserting a newly-arrived KV pair. The main steps of Circ-Tree in splitting are as follows.
\begin{enumerate}
	\item Circ-Tree first determines the split point for the LN, i.e., the middle position of the LN. 
	\item Circ-Tree allocates a new LN. From the split point, the greater half of KV pairs are copied into the newly-allocated LN.
	If the newly-arrived KV pair falls into the range of greater half, then we copy it alongside
	without shifting any KV pairs. 
	\item After copying, Circ-Tree fills the new LN's node header accordingly and sets the 
	sibling	pointer pointing to the right sibling of the original LN. 
	\item Circ-Tree alters sibling pointer pointing to the newly-allocated LN via an atomic write. It then clears copied KV pairs with $\mathsf{NULL}$s (zeros). If the newly-arrived KV pair falls into the range of smaller half, Circ-Tree inserts it into the original LN.
	\item Circ-Tree changes the number of keys in the original LN's node header via an atomic write, and starts updating upper-level INs where necessary. 
\end{enumerate}

Figure~\ref{fig:split} shows an example of splitting an LN by Circ-Tree.
Circ-Tree first finds the split point (\numbercircled{1}
in Figure~\ref{fig:split-a}). Then it allocates a new LN (\numbercircled{2}
in Figure~\ref{fig:split-b}).
We note that any newly-allocated node for Circ-Tree
is shredded with zeros ($\mathsf{NULL}$s)~\cite{NVM:shredder:ASPLOS-2016}.
Then Circ-Tree starts copying half KV pairs as well as the newly-arrived one
to the new LN, sets the base location and the number of keys as well
as the the sibling pointer pointing to the right sibling of the original LN  
with two successive atomic writes, respectively (\numbercircled{3} in Figure~\ref{fig:split-b}). Next, Circ-Tree alters the original LN's sibling
pointer to point to the new LN via an 8B atomic write, and nullifies  
 copied KV pairs in the original LN with $\mathsf{NULL}$s (\numbercircled{4}
in Figure~\ref{fig:split-b}). In the end, Circ-Tree updates the number of keys
in the original LN's node header with an 8B atomic write and starts updating
parental IN(s) where necessary (\numbercircled{5}
in Figure~\ref{fig:split-b}). This strict modification order is obeyed to  
sustain Circ-Tree's crash recoverability.

\subsection{Deletion}\label{sec:deletion}

The deletion with a key needs to locate a target LN as well as the exact position where
the KV pair is in the LN. 
In particular, Circ-Tree considers two cases when deleting a KV pair from an LN.
\begin{itemize}
	\item If the key to be deleted is the smallest or the greatest key of the LN, its value and key are cleared to be $\mathsf{NULL}$ with cache line flush and memory fence subsequently executed. 
	\item If the key under deletion falls inside the key range of the LN, Circ-Tree shifts KV pairs to the left or right. The shifting direction still depends on which direction shall incur fewer KV shifts.
\end{itemize}

In the first case,
using $\mathsf{NULL}$s to nullify the KV pair at either end of an LN helps to
set boundaries that can be leveraged to identify 
valid KV pairs in recovery.  
As to the second case, a removal of an in-between KV pair entails shifting KV pairs. 
If the KV pair to be deleted has a greater key than the middle KV pair,
Circ-Tree shifts greater KV pairs to the left and then sets
the KV pair of the greatest key at its original position to be $\mathsf{NULL}$s.
Otherwise, Circ-Tree shifts smaller KV pairs to the right and 
clears the KV pair of the smallest key at its original position using $\mathsf{NULL}$s.
Eventually 
Circ-Tree decreases the number of valid KV pairs by one
and/or updates the base location in the node header via an 8B atomic write.

\textbf{Merge}\hspace{2ex}
Continuous deletions dwindle the space of nodes.
A B+-tree node becomes underutilized
when its number of valid KV pairs drops below $\mathsf{N}/2$.
Circ-Tree considers
merging an underutilized node's KV pairs with its right sibling
if both of them are under the same parental IN
and the right sibling has sufficient vacant space to accommodate
all KV pairs of the underutilized node.
The reason why Circ-Tree only merges a node with its right sibling 
is for multi-threading, which will be explained later.
During a merge, KV pairs are first inserted into the right sibling.
Circ-Tree then atomically modifies the number of KV pairs and
the base location for the right sibling.
Then it resets the number of KV pairs in the underutilized node to be zero
via an atomic write. Next
the node linked list is adjusted
by detaching the underutilized node. 
Circ-Tree then updates upper-level INs where necessary.

In short, the procedures of merge is similar to that of split
and both of them must abide by strict modification orders instantiated in Figure~\ref{fig:split}
to avoid crash inconsistency.

\begin{figure*}[t] 
	\centering
	\begin{subfigure}[t]{0.7\columnwidth}
		\includegraphics[width=\textwidth]{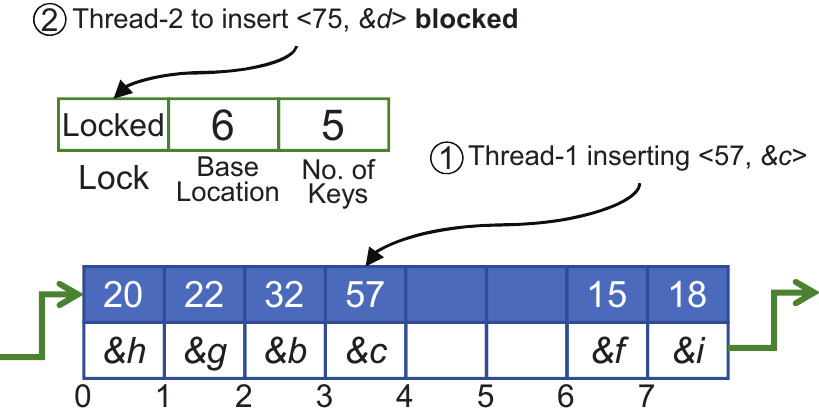}
		\caption{Multi-threading concurrent access to an LN}\label{fig:multi-a}
	\end{subfigure}
	\hfill
	\begin{subfigure}[t]{1.2\columnwidth}
		\includegraphics[width=\textwidth]{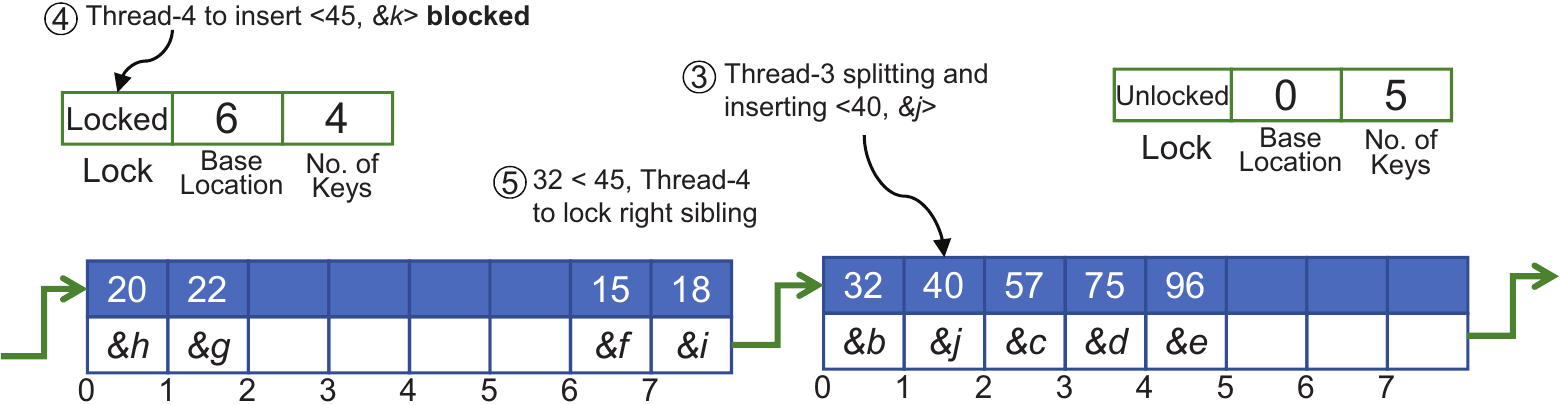}		
		\caption{Multi-threading concurrent access to an LN being split}\label{fig:multi-b}
	\end{subfigure}
	\vspace{-1ex}		
	\caption{An Example of Handling Multi-threading Accesses by Circ-Tree}~\label{fig:multi}
	\vspace{-3ex}
\end{figure*}

\subsection{Multi-threading Access}
As shown in Figure~\ref{fig:node},
each node header includes a lock  
to support multi-threading accesses.
Figure~\ref{fig:multi-a} captures two threads aiming to insert
KV pairs to the same LN. 
As Thread-1 has locked the LN and is doing
insertion (\numbercircled{1} in Figure~\ref{fig:multi-a}),
Thread-2 is being blocked (\numbercircled{2} in Figure~\ref{fig:multi-a}).

Figure~\ref{fig:multi-b} illustrates a more complicated scenario
as the current thread is splitting an LN that the other thread
has been waiting for (\numbercircled{3} and \numbercircled{4} in Figure~\ref{fig:multi-b}).
To avoid a KV pair to be incorrectly inserted, 
Circ-Tree checks whether the key under insertion is greater than the smallest one in the LN's right sibling. 
If so, the current thread releases the lock but locks the right sibling (\numbercircled{5} in Figure~\ref{fig:multi-b}).
Otherwise, it proceeds with the insertion.

Parental IN may need to be split with multi-threading access.
In this scenario, before releasing the LN's lock, the current thread locks and operates the parental IN.
It may further lock upper-level INs up to the root. If an IN is being locked by another thread,
the current thread will wait and may need to traverse to the IN's right sibling to proceed. Once the thread
finishes an insertion into the highest IN where necessary, it unlocks the IN and goes down to unlock lower-level
INs and the LNs. In other words, Circ-Tree recursively locks and unlocks an LN and its parental INs.

Circ-Tree deals with a merge under multiple threads similarly to a split
When a thread is merging KV pairs of an underutilized LN
to its right sibling, the other insertion/deletion/search thread 
that waits for the lock of the underutilized LN
later finds the number of KV pairs changes to be zero.
Similarly, this thread will go to the right sibling to continue.
In the multi-threading environment,
the sibling pointer of 
an underutilized node that has been merged is not immediately
cleared if the lock in the node header is on hold.

\subsection{Recovery of Circ-Tree}\label{sec:recover}
The recoverability of Circ-Tree is entitled by 1) the 8B 
atomic update of the base location and the number of valid keys as well as two pointers in the node header,
2) the $\mathsf{NULL}$ boundaries and non-duplicate valid values
in the array of KV pairs, and 3) strict execution orders of insertion 
and deletion.
The property of non-duplicate values in a B+-tree node has been
exploited~\cite{NVM:FAST+FAIR:FAST-2018}. However, as Circ-Tree bidirectionally shifts KV pairs,
it needs further efforts for crash recovery.
To detect the occurrence of crashes, 
a special flag is installed in the root node of Circ-Tree. It is
flagged up when Circ-Tree starts and cleared in case of a normal exit. 
If a crash happens, Circ-Tree will be conscious of the uncleared flag 
and initiate a recovery.
It traverses all tree nodes in a bottom-up way, i.e., from LNs to INs.
Circ-Tree scans successive LNs to discover possible inconsistency scenarios.
\begin{enumerate}[leftmargin=*]
	\item The number of keys is one smaller than the non-$\mathsf{NULL}$ values in the array of KV pairs. This is caused by an incomplete insertion. 
	\begin{enumerate}[leftmargin=*]
		\item If there are duplicate non-$\mathsf{NULL}$ values,  
		which means the crash occurred during shifting KV pairs, Circ-Tree will use the base location to
		decide how to undo the insertion. In brief, if the value at the base location's logical left is non-$\mathsf{NULL}$, which means shifting to the left was being performed prior to the crash, 
		Circ-Tree shifts to the right from the logical leftmost KV pair until the duplicate value is shifted back; otherwise, Circ-Tree shifts KV pairs to the left until the duplicate value is shifted back.
		\item If no duplicate non-$\mathsf{NULL}$ values exist, that means shifting KV pairs had been completed but the crash happened before atomically updating the base 
		location and the number of KV pairs. Circ-Tree atomically modifies them 
		to bring the LN to a consistent state.
	\end{enumerate}
	\item The number of keys is one greater than the non-$\mathsf{NULL}$ values in the array of KV pairs. 
	This is caused by an incomplete deletion. Similarly to aforementioned cases with an incomplete insertion, 
	Circ-Tree uses the number of non-$\mathsf{NULL}$ values, the existence of duplicate values and the base location to fix the inconsistency.
	\item Two sibling LNs contain duplicate valid values. That means a crash has happened before the completion of a split or merge procedure.
	\begin{enumerate}[leftmargin=*]
		\item If both LNs are consistent concerning their respective arrays and node headers, the crash must have happened 1) before
		the two LNs' parental IN was updated in split, 
		since updating the LN's node header and parental IN is the last step 
		in split (cf.~\numbercircled{5} in Figure~\ref{fig:split} for split), and 
		2) before the number of KV pairs was reset to be zero in merge.
		By analyzing these two LNs and their parental IN, 
		Circ-Tree can proceed and complete the split or merge.
		\item If the number of non-$\mathsf{NULL}$ KV pairs in one LN is not the same as the number of keys in this LN's node header, 
		that means either setting $\mathsf{NULL}$s has not been done in splitting this LN (cf.~\numbercircled{4} in Figure~\ref{fig:split}),
		or resetting the number of keys as zero in
		the LN's node header during merge has been done.
		Circ-Tree fixes such inconsistencies by removing duplicate KV pairs from the LN's array and updating its node header. It might update the LN's parental IN if the other LN is not recorded in the parental IN during split.
	\end{enumerate}	
\end{enumerate}

After scanning LNs, Circ-Tree scans and fixes inconsistent INs 
with similar strategies in addition to a double check of each IN's 
	children nodes. Because a crash may occur at the last step of split 
	(resp. merge), i.e., just before \numbercircled{5} in Figure~\ref{fig:split-b}, 
	the number of pointers stored in an IN can be one less (resp. more) than the
	actual children nodes although no duplicate pointers exist in the IN. 
	Given such an IN, 
	Circ-Tree traverses its corresponding children in the next-level node linked list and compares their addresses to the pointers recorded in the IN. 
	If one child is found to be missing (resp. present) in the IN, Circ-Tree inserts (resp. deletes) 
	it within the IN. Circ-Tree does so up to the root in a  
	bottom-up fashion.
In recovery, Circ-Tree also reinitializes all locks in node headers
to be unlocked.
In addition, when using Circ-Tree to build a KV store system,
we demand that, to insert a KV pair with concrete value, the value itself must be persisted into NVM 
before inserting the key and the pointer to the committed value into Circ-Tree. 
Modifying a value is conducted in the copy-on-write way
before replacing the corresponding pointer in the Circ-Tree.
\section{Evaluation}~\label{sec:evaluation}

In this section, we first evaluate Circ-Tree 
as a standalone B+-tree variant by comparing to other volatile
or in-NVM B+-tree variants
under single-threading
and multi-threading workloads.
Then we test the efficacy of Circ-Tree when it is used
as the indexing structure in a KV store system. To do so, we employ the prevalent YCSB~\cite{YCSB} for benchmarking.

\subsection{Evaluation Setup}

\textbf{Platform}\hspace{2ex}
We have used a Dell OptiPlex 7050 in which there is an Intel Core$^{\textnormal{TM}}$ 
i7-7700 CPU with 256KB/1MB/8MB for L1/L2/L3 caches, respectively.
The CPU provides {\tt clflushopt} for cache 
line flush. The instruction used for memory fence is 
{\tt sfence}. The operating system is Ubuntu 18.04.2 and the compiler version is GCC/G++ 7.3.1.

The machine has a 64GB DRAM and
we use a part of it to emulate the NVM space. 
NVM technologies generally have asymmetrical write/read latencies. 
We keep the read latency of NVM the same as that of DRAM, and emulate 
the write latency by adding an extra delay after each {\tt clflushopt} 
instruction~\cite{NVM:NVWAL:ASPLOS-2016, kv:hash:OSDI-2018}.
Following previous works~\cite{NVM:FAST+FAIR:FAST-2018, kv:NoveLSM:ATC-2018}, 
we set the default write latency of NVM as 300ns. 

\begin{figure*}[t] 
	\centering
	\begin{subfigure}[t]{0.485\columnwidth}
		\adjincludegraphics[valign=t, width=\textwidth]{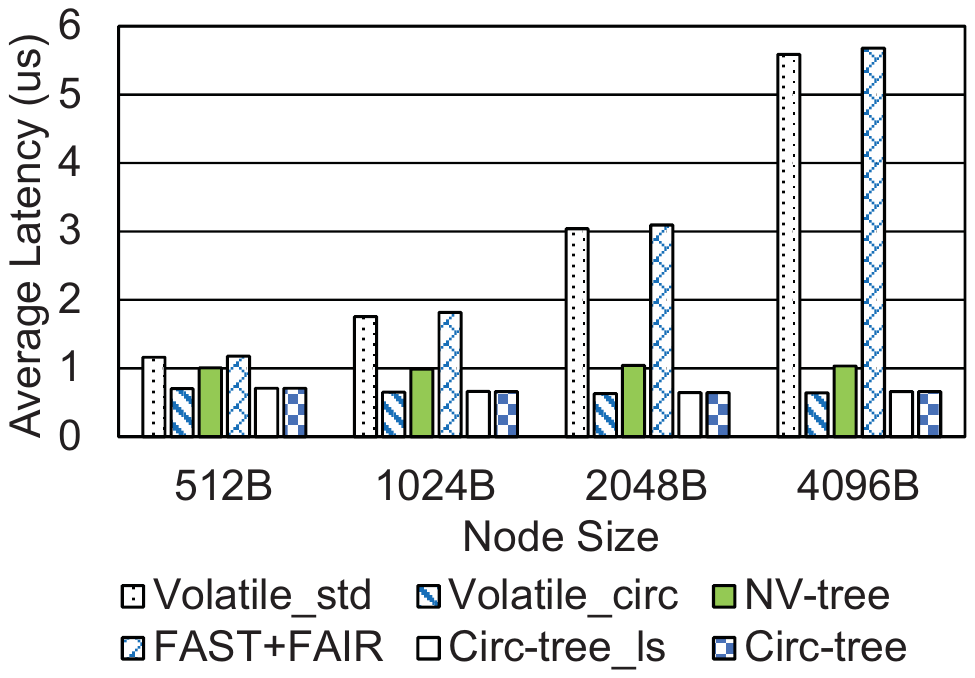}
		\caption{Average insert latency ($\mu$s)}
		\label{fig:single-avg-latency}
	\end{subfigure}
	\hfill
	\begin{subfigure}[t]{0.485\columnwidth}
		\adjincludegraphics[valign=t, width=\textwidth]{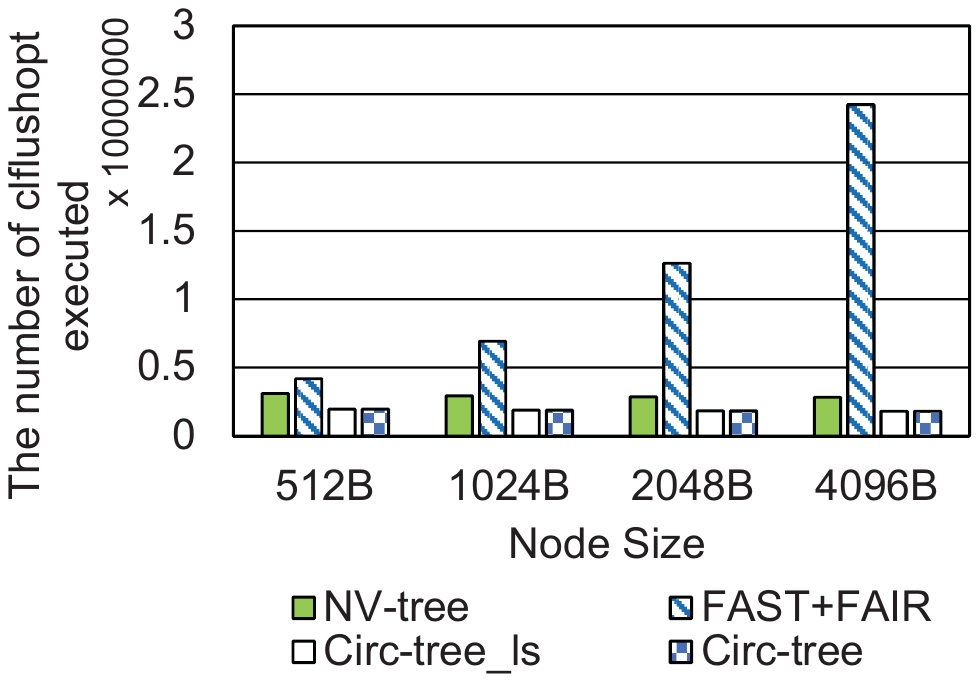}			
		\caption{The number of {\tt clflushopt}}\label{fig:single-clflushopt}
	\end{subfigure}	
	\hfill
	\begin{subfigure}[t]{0.485\columnwidth}
		\adjincludegraphics[valign=t, width=\textwidth]{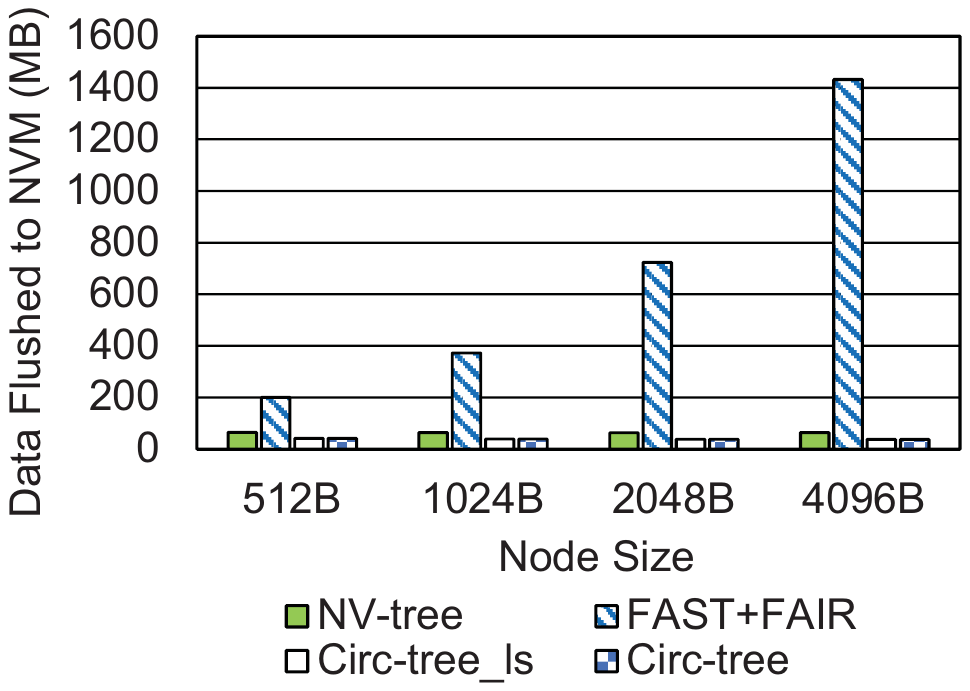}				
		\caption{The amount of flushed data}\label{fig:single-data-flushed}
	\end{subfigure}	
	\hfill
	\begin{subfigure}[t]{0.495\columnwidth}
		\adjincludegraphics[valign=t, width=\textwidth]{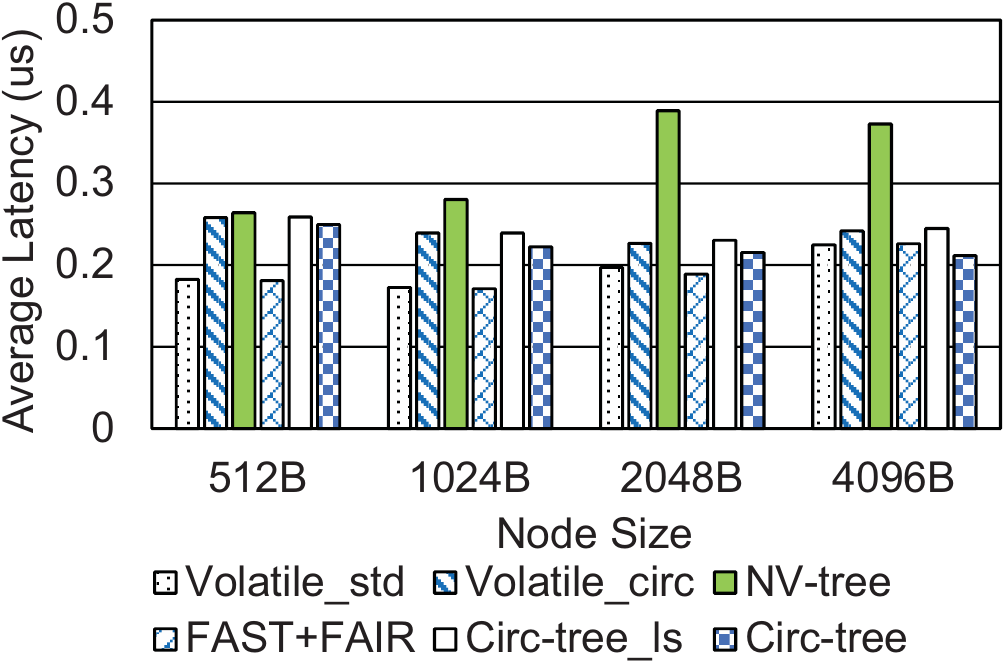}	
		\caption{Average latency (search, $\mu$s)}\label{fig:single-search-latency}
	\end{subfigure}		
\vspace{-1ex}	
	\caption{A Comparison of Six Trees on Inserting and Searching 1 Million Keys}~\label{fig:single-write}
\vspace{-4ex}		
\end{figure*}

\textbf{Competitors}\hspace{2ex}
We have used C++ to implement two versions of Circ-Tree. 
One is a Circ-Tree with circular node structure and linear search starting 
from the base location of a node to the last valid KV pair (Circ-Tree\_ls).
The other Circ-Tree
performs linear search only in a contiguous space 
of a circular node.  .
We employ other four 
B+-tree variants to compare against Circ-Tree. 
We implemented a standard volatile B+-tree (Volatile\_std)
without forcefully flushing any data to NVM (without {\tt clflushopt} or {\tt sfence}
but retaining the write latency). We also implemented another volatile B+-tree 
(Volatile\_circ),
the node of which is in the circular fashion. 
NV-tree and FAST+FAIR are two state-of-the-art in-NVM B+-tree variants  
that represent two approaches with unsorted and sorted tree nodes, respectively.
FAST+FAIR has open-source code\footnote{Available at {https://github.com/DICL/FAST\_FAIR}.} 
 while we implemented NV-tree 
in line with its original literature~\cite{NVM:NV-tree:FAST-2015}.
All implementations have been compiled with -O3 option.

We used 512B, 1KB, 2KB, and 4KB for the node sizes.
We note that the node size refers to the size of the array of KV pairs.
For a fair comparison, in any node of each tree,
we made the array of KV pairs cache line-aligned and also separated
the node header into an individual cache line.

For end-to-end comparison, we built four KV store prototypes with Circ-Tree, Circ-Tree\_ls, NV-tree, and FAST+FAIR, respectively. These KV store systems include interfaces to receive and handle access requests issued by YCSB. The key from a YCSB workload is 
a string with a prefix (`{\em user}') 
and a number.
We removed the prefix and used an unsigned 8B integer to treat 
the number as a key. The default value of YCSB has ten fields with 100B per field, so
we made a two-dimensional (2D) array for each value. The first dimension holds pointers to ten fields.
To add a new key-value, 
the actual value must be committed and flushed to NVM before 
inserting its KV pair in 
an indexing tree.  
To update one field of a value, we used copy-on-write to first write the field
elsewhere and change the field's pointer in the value's 2D array.

\textbf{Workloads}\hspace{2ex}
 To evaluate standalone B+-tree variants, we followed
the widely-used uniform distribution~\cite{NVM:NV-tree:FAST-2015, NVM:FPTree:SIGMOD-2016, NVM:HiKV:ATC-2017, NVM:FAST+FAIR:FAST-2018} to generate one million, ten million, and 100 million unsigned 64-bit keys that do not exhibit 
access skewness. Each key was inserted with an 8B pointer as a value into a tree.
We used the SessionStore (numbered as `workloada') workload with YCSB. 
It first inserts a predefined number of KV pairs and then follows a search/update 
ratio of 50\%/50\% over keys selected in accordance with a Zipf distribution. 
The number of inserted KV pairs in our experiment is 1 million per client thread 
while the total number of search and update requests is also 1 million (0.5/0.5 million) 
per client thread.
In addition, deletion is just a reverse operation of insertion for B+-tree
by shifting KV pairs in opposite directions, so the difference among deletion performances
of trees is similar to that of insertion performances.
Due to the space limitation,
we focus on showing the performances of insertion and search.

\textbf{Metrics}\hspace{2ex}
We use the 
average latency per insertion/search
 as the main metric to measure performance. A shorter average latency means a higher performance.
However, using arithmetic mean 
to calculate average latency 
is inaccurate due to its biases to very short or very long latencies~\cite{bench:summary:ACM-1986}.  
For a standard B+-tree, inserting a new 
greatest key into a node takes significantly shorter time than inserting 
a key that incurs splits up to the root.
As a result, we have chosen  
the geometric mean 
to calculate the average
latency~\cite{bench:summary:ACM-1986,bench:crimes}. 
For end-to-end comparison, 
YCSB reports a series of latencies 
from which we choose the 99th percentile latency to rule out the impact of very short or very long operations.
It means that 99\% of overall write/read requests 
can be completed below such a latency.
In addition, each number of the results presented in following subsections is the average value
 in geometric mean by running the respective experiment for five times.

\subsection{Performance Comparisons on Insertion and Search}

Figure~\ref{fig:single-avg-latency}  
captures the average latencies  
of six trees on inserting one million keys with four node sizes.
From Figure~\ref{fig:single-avg-latency}, we first observe that Circ-Tree 
substantially outperforms NV-tree and FAST+FAIR with much shorter latencies.
For example, with 4KB node, 
the average latencies of NV-tree and FAST+FAIR are 1.6$\times$ and 8.6$\times$
that of Circ-Tree. 
We have recorded the number of {\tt clflushopt} executed 
by NV-tree, FAST+FAIR, Circ-Tree\_ls, and CIRC-Tree to
flush data to NVM and the overall amount of flushed data.
These two numbers help to explain the reason for Circ-Tree to yield 
much higher write performance.
As indicated by Figure~\ref{fig:single-clflushopt} and Figure~\ref{fig:single-data-flushed},
Circ-Tree incurred the fewest {\tt clflushopt} and flushed the least amount of data.
With 4KB node size, the numbers of executed {\tt clflushopt} of NV-tree and FAST+FAIR
are 1.6$\times$ and 13.3$\times$ that of Circ-Tree, respectively, while the amount of data flushed
by NV-tree and FAST+FAIR are 1.7$\times$ and 37.8$\times$ that of Circ-Tree, respectively.
The substantial reductions of executed {\tt clflushopt} and flushed data in turn
justify the reduction of write amplifications achieved by Circ-Tree.

Let us first make a comparative analysis between Circ-Tree and FAST+FAIR.
FAST+FAIR keeps keys sorted in a linear node structure and shifts KV pairs in a
unidirectional way. Circ-Tree employs a circular node and bidirectionally
shifts KV pairs. In practice,
the impact of linear structure and unidirectional shifting is cascading and profound. 
Let us consider a special case in which we continue to insert a new smallest key
to a 4KB linear node until the node is full. A 4KB node can hold at most 256 KV pairs. From inserting the second KV pair, all existing 
KV pairs must be shifted. Therefore, in all,
$(0 + 1 + 2 + \cdots + 255) = 32,640$ KV pairs have to be shifted. Nonetheless, for 
a circular node, no shift is required for any insertion.
Concretely, Circ-Tree surely conducts much fewer memory writes to NVM than FAST+FAIR
and in turn significantly reduces write amplifications.

The reason why Circ-Tree outperforms NV-tree is multifold.
First, although NV-tree uses the append-only fashion to avoid shifting KV pairs,
the way it splits an LN is time-consuming. This is because it 
must first scan all unsorted KV pairs
to separate smaller half from greater half, and then copy and flush
them into two newly-allocated LNs. 
Secondly, NV-tree needs a flag for each KV pair to label the validity 
of the KV pair, which decreases space 
utilization of an LN and holds fewer KV pairs than FAST+FAIR and Circ-Tree.
Hence, given the same insertion workload, more splits are expected for NV-tree.
These explain why NV-tree 
flushed more data than FAST+FAIR and Circ-Tree.
Thirdly, the design of NV-tree demands that it has to stall and
rebuild INs even though only one IN becomes full. This is because NV-tree organizes all INs
in a contiguous memory space which cannot be adjusted except  
a rebuilding~\cite{NVM:NV-tree:FAST-2015}.
Finally, every insertion for NV-tree has to traverse all KV pairs stored 
in an unsorted node to find out whether a previous valid version exists. 

\begin{figure}[t]		
	\centering
	\begin{subfigure}[t]{0.49\columnwidth}
		\includegraphics[width=\textwidth]{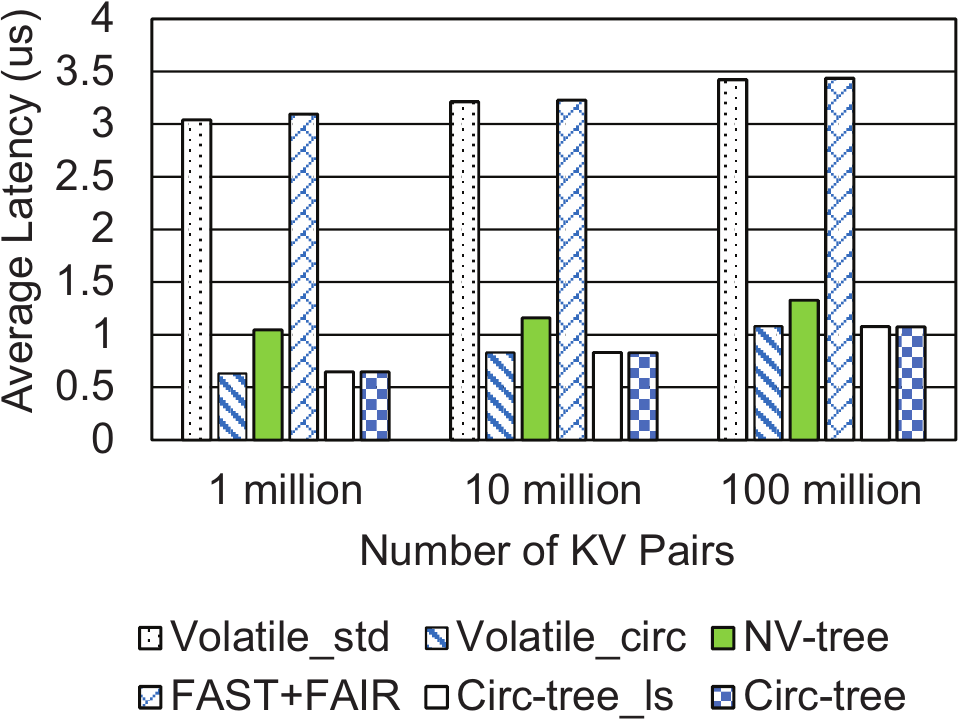}
		\caption{Average latency (insert, $\mu$s)}\label{fig:10m-100m}
	\end{subfigure}
	\hfill
	\begin{subfigure}[t]{0.49\columnwidth}
		\includegraphics[width=\textwidth]{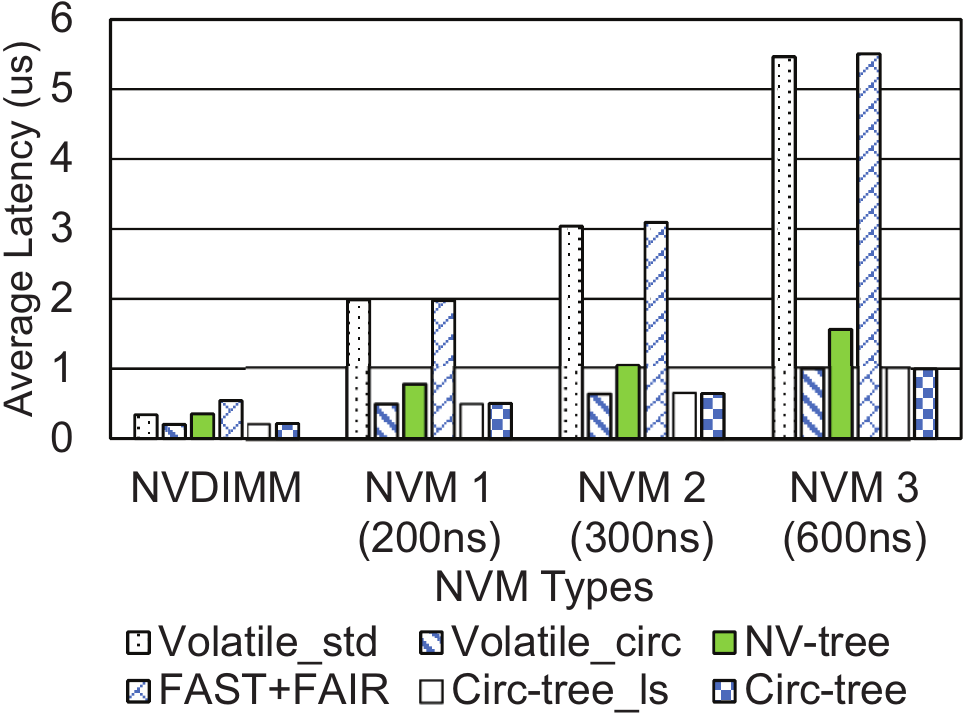}		
		\caption{Average latency (insert, $\mu$s)}\label{fig:nvm}
	\end{subfigure}
\vspace{-1ex}	
	\caption{A Comparison of Six Trees on (a) Inserting 1/10/100 Million KV Pairs, and (b) Three NVM Types}~\label{fig:discussion}	
	\vspace{-3ex}	
\end{figure}

From Figure~\ref{fig:single-avg-latency}, 
we also observe that the performance gap between FAST+FAIR and Circ-Tree becomes wider with a larger node size.
From 512B to 4KB, the average latency of FAST+FAIR is 1.7$\times$, 2.7$\times$, 4.8$\times$,
and 8.6$\times$ that of Circ-Tree, respectively. 
A smaller node holds fewer KV pairs. 
Hence, the performance gain brought by bidirectional shifting decreases.
However, given the same amount of data stored, a smaller node size
entails much more INs for indexing. 
For instance, assume that there are $2^{20}\approx$ 1 million keys that
are already ordered, so they can be 
densely stored in B+-tree nodes. Given 512B node that holds at most 32 KV pairs, 
the height of a B+-tree is four with one, 32, and 1024 INs at three upper levels, respectively.
These INs take about $(1 + 32 + 1024)\times 512 \approx 529$KB NVM space.
As to the 4KB node that holds a maximum of 256 KV pairs,
the height of B+-tree is three with one and 16 INs at two upper levels, respectively.
These INs cost $(1 + 17)\times 4096 = $68KB NVM space.   
Therefore, 512B node demands $\frac{529}{68}\approx$ 7.8$\times$ NVM space for INs compared 
to 4KB node. Besides, more INs demand extensive memory allocations and links 
of nodes into a tree.
We note that INs are only for indexing and can be always 
reconstructed from LNs~\cite{NVM:NV-tree:FAST-2015}.
As a result, Circ-Tree prefers a larger node size.

The third observation illustrated by Figure~\ref{fig:single-avg-latency}
is that the performance of Circ-Tree remains consistent with variable node sizes.
Circ-Tree  
shifts KV pairs bidirectionally and is hence not affected by the node size.
Because of the slow write speed of NVM, the average latency of
Circ-Tree is comparable to that of Volatile\_Circ.
Both Volatile\_std and FAST+FAIR
unidirectionally shift KV pairs to keep keys sorted, so
a larger node size necessitates more KV pairs to be shifted. Hence, 
their average latencies rise up with larger node sizes.

The fourth observation with Figure~\ref{fig:single-avg-latency} is that 
the average latency for insertion of Circ-Tree\_ls is similar to that of Circ-Tree.
The reason is that the two only differ in searching but
shifting and writing KV pairs to NVM is the dominant factor affecting the insertion performance.

Figure~\ref{fig:single-search-latency} captures the average latency 
 of six trees in searching
for one million inserted KV pairs with four node sizes.
Owing to unsorted KV pairs in LNs, NV-tree yields the worst search performance.
The average latency of Circ-Tree is shorter than Volatile\_circ and Circ-Tree\_ls because 
the former only searches in a contiguous memory space and avoids jumping between disjointed
cache lines. 

Moreover, 
with smaller node sizes, Circ-Tree is slower than Volatile\_std and 
FAST+FAIR. A smaller node size helps them 
to maintain cache efficiency by prefetching. However,
searching a smaller node 
or a part of it by Circ-Tree does not differ much. 
This explains with 512B node, even the average latency of NV-tree is 
close to that of Circ-Tree.
Circ-Tree calls conditional
branches to speculate about whether KV pairs are contiguous 
and which segment it should search.
The cost of such branches offsets the gain introduced by searching
a contiguous part of a node. However,
with a larger node size, e.g., 4KB, 
the average latency for search of Circ-Tree is 7.1\% shorter than 
that of Volatile\_std and FAST+FAIR.

\subsection{Impact of Workload and NVM}

We also tested inserting ten million and 100 millions keys 
in uniform distributions with 2KB node.
Figure~\ref{fig:10m-100m} captures the average latencies for six trees.
Circ-Tree still achieved the highest performance with heavier workloads. Nonetheless,
all trees have longer average latency when the workload increases.
For example, from ten million to 100 million, the average latencies of FAST+FAIR and Circ-Tree
increase by 6.5\% and 29.5\%, respectively.
With increasing number of keys stored in a B+-tree variant,
inserting new keys to the tree becomes more time-consuming.
The reason for such increased overhead is twofold.
First, more keys lead to the increase of
the height of a B+-tree variant. Thus, the traversal time from the tree root
to a target LN turns to be longer, because 1)
more cache misses occur due to more nodes to be loaded and searched, and 2)
more comparisons are performed to locate the appropriate node at the next levels.
Secondly, splits that spread to upper levels may involve more INs.

\begin{figure}[t]		
	\centering
	\begin{subfigure}[t]{0.49\columnwidth}
		\includegraphics[width=\textwidth]{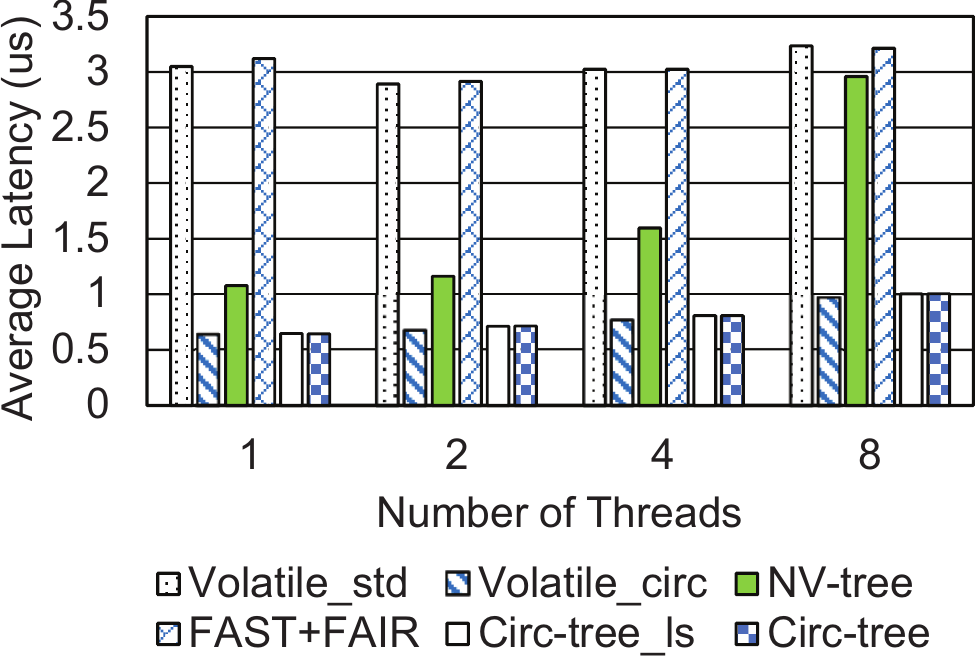}
		\caption{Average latency (insert, $\mu$s)}\label{fig:multi-insert-avg}
	\end{subfigure}
	\begin{subfigure}[t]{0.49\columnwidth}
		\includegraphics[width=\textwidth]{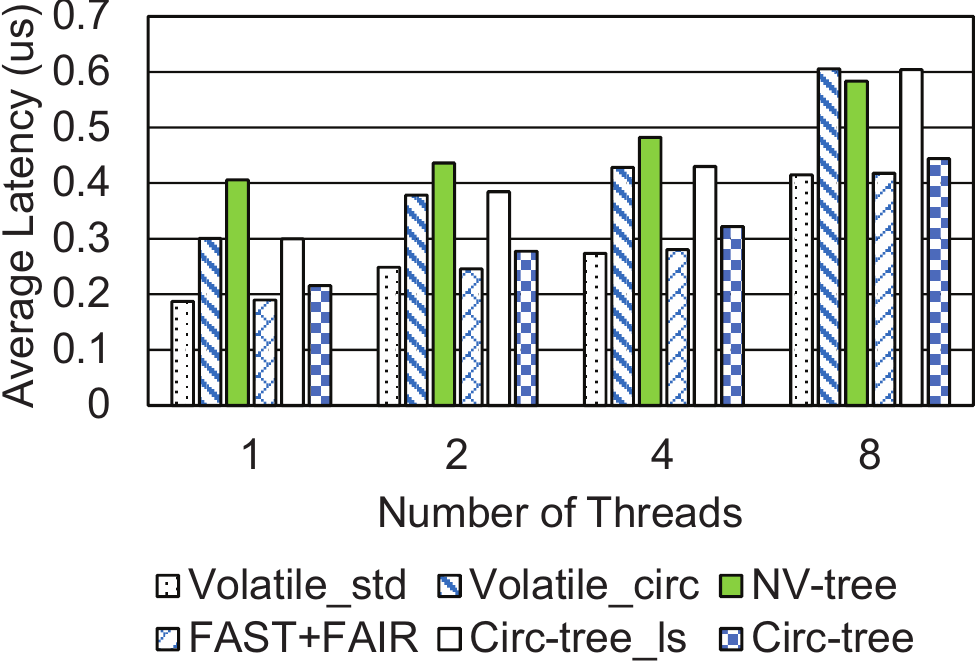}	
		\caption{Average latency (search, $\mu$s)}\label{fig:multi-search-avg}
	\end{subfigure}
\vspace{-1ex}	
	\caption{A Comparison of Six Trees with Multi-threading}~\label{fig:multi-threading}
\vspace{-3ex}		
\end{figure}

We used 300ns as the default write latency for NVM in experiments. Researchers have used NVDIMM with the same write speed as DRAM~\cite{NVM:NV-tree:FAST-2015,NVM:soft-updates:ATC-2017}, and NVM with 200ns and 600ns write latencies~\cite{NVM:HiKV:ATC-2017,NVM:soft-updates:ATC-2017,kv:hash:OSDI-2018}.
We also considered these three configurations in enumlating NVM and did experiments of inserting
one million keys with 2KB node.
Figure~\ref{fig:nvm} shows the average latencies of six trees with
four NVM types. 
With slower NVM, the cost of writing data to NVM is higher, so the write
amplifications caused by shifting KV pairs are more significant.
This explains why the performance gaps between Circ-Tree and FAST+FAIR are
2.5$\times$, 3.9$\times$, 4.8$\times$, and 5.5$\times$ with increasing write latencies.

\begin{figure*}[t]		
	\centering
	\begin{subfigure}[t]{0.65\columnwidth}
		\includegraphics[width=\textwidth]{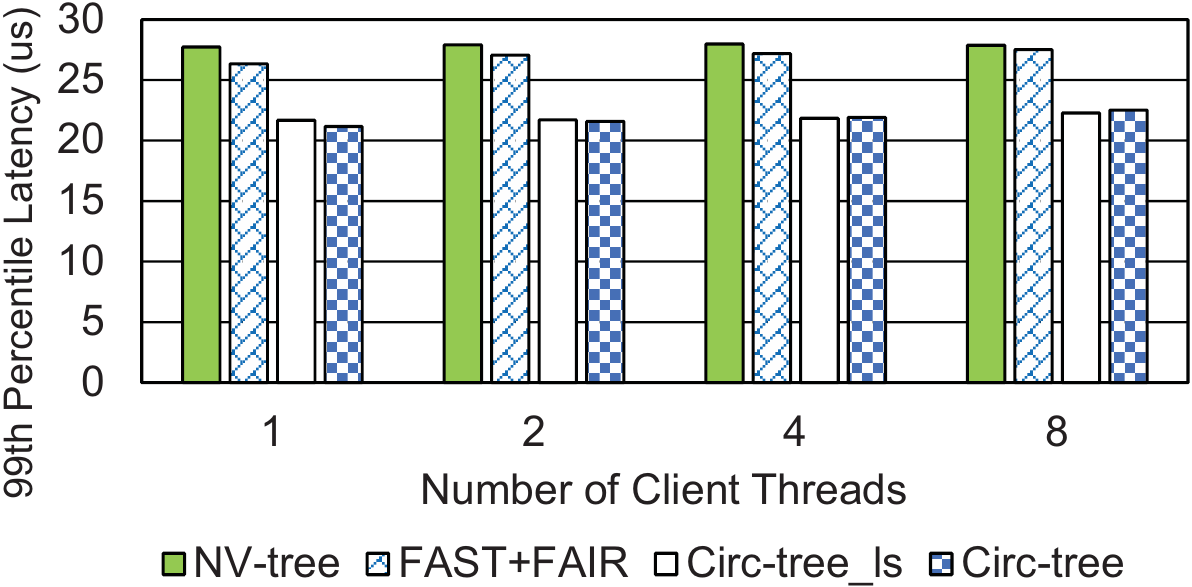}
		\caption{99th percentile latency (insertion, $\mu$s)}\label{fig:kv-2KB-insert}
	\end{subfigure}
	\hfill
	\begin{subfigure}[t]{0.65\columnwidth}
		\includegraphics[width=\textwidth]{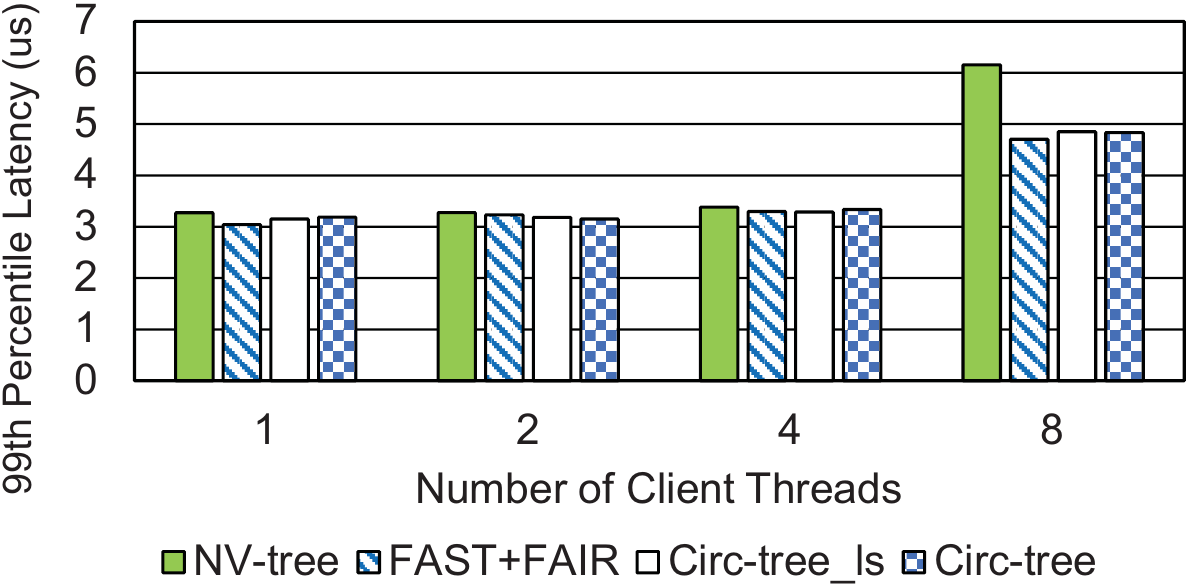}		
		\caption{99th percentile latency (search, $\mu$s)}\label{fig:kv-2KB-search}
	\end{subfigure}
	\hfill
	\begin{subfigure}[t]{0.65\columnwidth}
		\includegraphics[width=\textwidth]{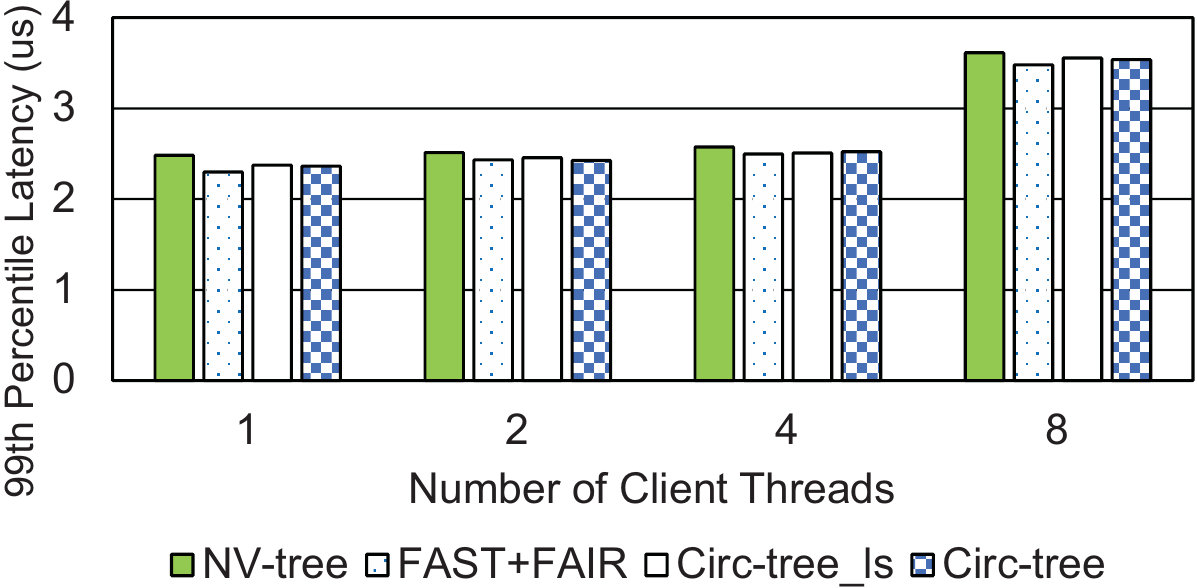}		
		\caption{99th percentile latency (update, $\mu$s)}\label{fig:kv-2KB-update}
	\end{subfigure}
\vspace{-1ex}	
	\caption{A Comparison of Four KV Stores (2KB Node) on SessionStore Worload of YCSB}~\label{fig:kv-2KB}
\vspace{-3ex}		
\end{figure*}

\begin{figure*}[t]		
	\centering
	\begin{subfigure}[t]{0.65\columnwidth}
		\includegraphics[width=\textwidth]{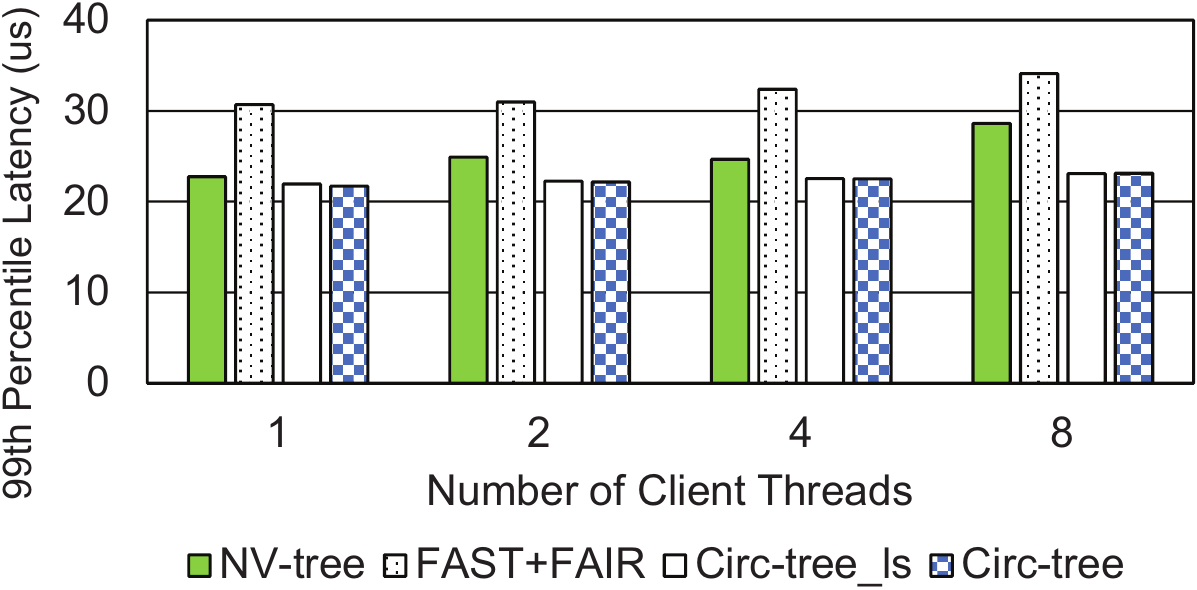}
		\caption{99th percentile latency (insertion, $\mu$s)}\label{fig:kv-4KB-insert}
	\end{subfigure}
	\hfill
	\begin{subfigure}[t]{0.65\columnwidth}
		\includegraphics[width=\textwidth]{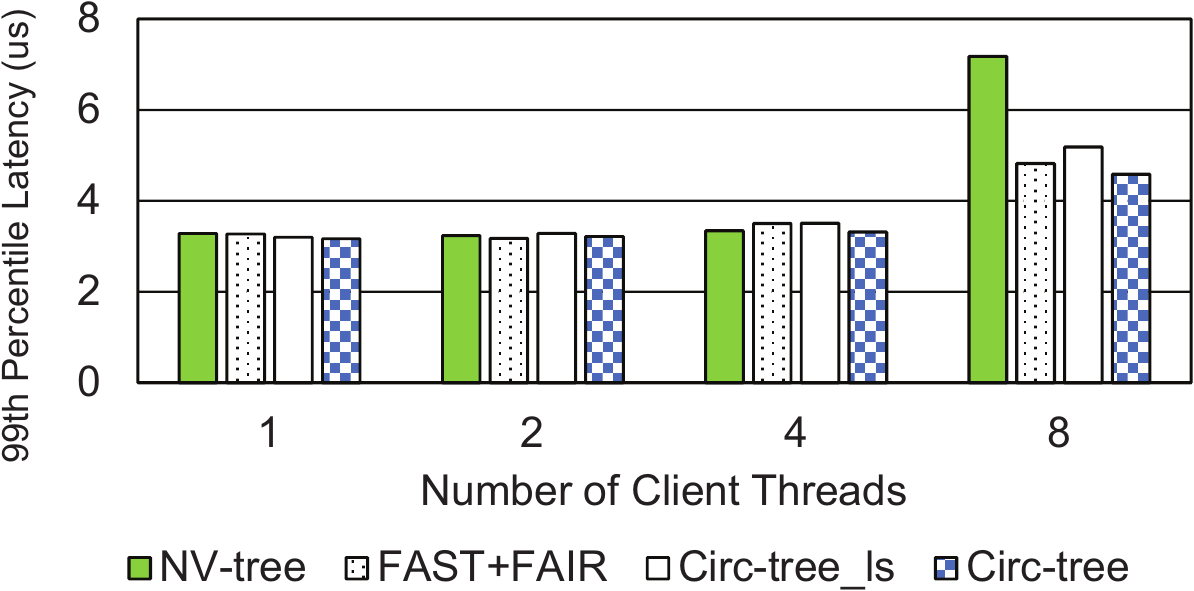}		
		\caption{99th percentile latency (search, $\mu$s)}\label{fig:kv-4KB-search}
	\end{subfigure}
	\hfill
	\begin{subfigure}[t]{0.65\columnwidth}
		\includegraphics[width=\textwidth]{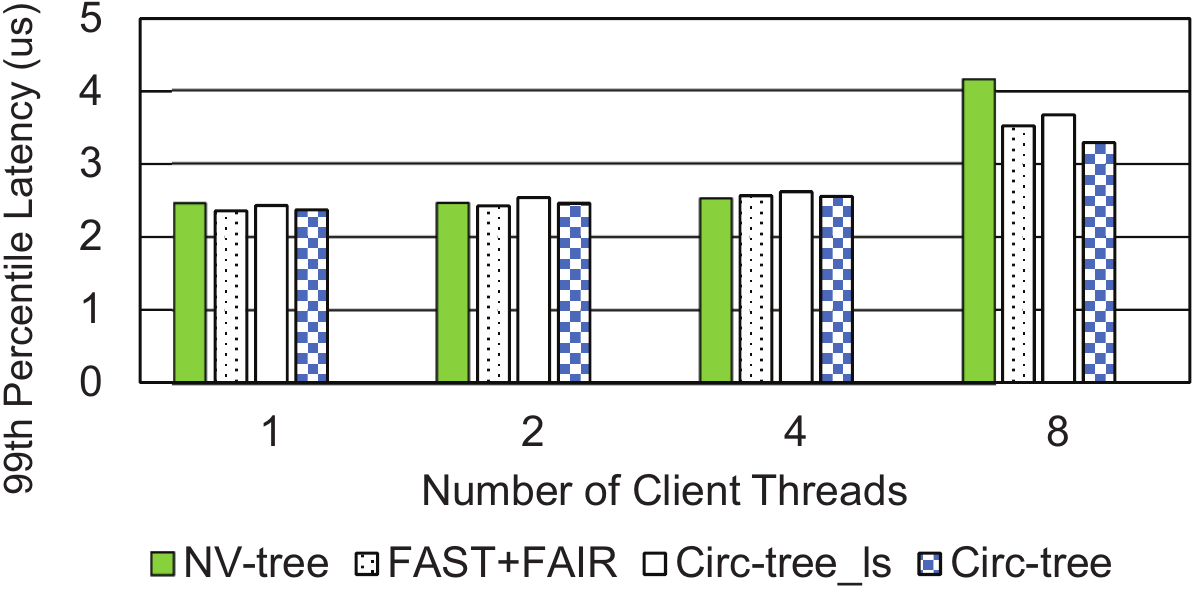}		
		\caption{99th percentile latency (update, $\mu$s)}\label{fig:kv-4KB-update}
	\end{subfigure}
\vspace{-1ex}	
	\caption{A Comparison of Four KV Stores (4KB Node) on SessionStore Worload of YCSB}~\label{fig:kv-4KB}
\vspace{-3ex}		
\end{figure*}

\subsection{Multi-threading Performance}

We have performed multi-threading tests to evaluate the performance of 
Circ-Tree. We varied the number of threads to be 1, 2, 4, and 8.
We generated eight sets, each of which has one million keys
in the uniform distribution. There are no duplicate keys across these 
sets. Each thread inserted and searched with a set.
We used 2KB as the node size for every tree.
Figure~\ref{fig:multi-threading} shows 
the average latencies
for insertion and search with varying the number 
of threads for six trees.
Circ-Tree and Circ-Tree\_ls
still significantly outperform NV-tree and FAST+FAIR even in the 
presence of multiple threads.  
For example, with eight threads, the average latencies for insertion of NV-tree and FAST+FAIR
are 2.9$\times$ and 3.2$\times$ that of Circ-Tree, respectively.

With more threads,  
B+-tree variants require more time to handle concurrent
insertion requests.  
One reason is similar to the one mentioned for inserting
1/10/100 million keys, as more threads gradually
make more KV pairs inserted. Moreover, when more threads 
concurrently insert KV pairs, the lock period due to 
operating with the same nodes becomes longer. Trees 
that are more efficient in the context of single-threading 
insertion, suffer more from a longer lock period (i.e. the 
duration a tree node might be locked for insertion or search).
For instance, the average latencies for insertion of FAST+FAIR 
and Circ-Tree increased by 6.0\% and
19.5\%, respectively, when the number of operating threads increases 
from four to eight.  
Since Circ-Tree leverages bidirectional shifting to
reduce write amplifications, the insertion performance 
of Circ-Tree is impaired more by the longer lock period.
For NV-tree, however, its average latency almost doubled 
when the number of operating threads was increased from 
four to eight. 
Specifically, when more threads are inserting KV pairs, INs of NV-tree are more likely
to become full. A full IN entails a rebuilding of all INs, which
blocks all insertion threads. 
With more threads inserting data, NV-tree stalls more.

As shown in Figure~\ref{fig:multi-search-avg}, more threads also lead to 
longer average latencies for search as they contend to access the same nodes. 
Though, both FAST+FAIR and Circ-Tree suffer from
the longer lock period with more threads, so
the performance gap between them decreases
from 11.6\% to 5.9\% with four and eight threads, respectively. 
In addition, more threads also confirm the capability of
Circ-Tree's search in a contiguous space. A comparison
between Circ-Tree\_ls and Circ-Tree with 8 threads shows
that, the average latency for search of Circ-Tree is 36.1\% shorter
than that of Circ-Tree\_ls.

\subsection{Recovery Time}

We emulated inconsistency scenarios because the NVM used in evaluation was based on volatile DRAM.
We first inserted 1/10/100 million keys.
Then we orderly saved all INs and LNs into a text file. When saving nodes into the file,
we randomly selected some nodes and either decreased the number of keys by one or
shifted KV pairs to generate duplicate non-$\mathsf{NULL}$ values.
Next we reconstructed the tree by allocating, filling, and linking nodes.
In the final step, we called the recovery procedure.
For each number of keys (1/10/100 million), 
we repeated the recovery test for five times, and
all inconsistency issues were found and fixed. 
Table~\ref{tab:recovery} shows the average recovery time calculated in geometric mean.
With a smaller node size, the recovery spent more time because much more INs were
processed. With the 4KB node, the recovery time of Circ-Tree for 100 million KV pairs
is just 352.7$m$s (0.35 second) with NVM having the same read latency as DRAM.
We believe this recovery time is acceptable in practice.

\begin{table} [t]
	\centering
	\caption{The Recovery Time of Circ-Tree (unit: $m$s)}\label{tab:recovery}	
	\vspace{-1ex}	
	\scriptsize
	\begin{tabularx}{\columnwidth}{|X|X|X|X|X|}		
		\hline
		\multicolumn{1}{|c|}{Number of KV}  & \multicolumn{4}{c|}{Node Size}\\ \cline{2-5}
		\multicolumn{1}{|c|}{Pairs Stored} & \multicolumn{1}{c|}{512B} & \multicolumn{1}{c|}{1KB} & \multicolumn{1}{c|}{2KB} & \multicolumn{1}{c|}{4KB}\\ \hline
		\multicolumn{1}{|r|}{1 Million} & \multicolumn{1}{r|}{\ \ 14.9} & \multicolumn{1}{r|}{\ \ 8.8} & \multicolumn{1}{r|}{\ \ 5.1} & \multicolumn{1}{r|}{\ \ 3.1} \\ \hline
		\multicolumn{1}{|r|}{10 Million} & \multicolumn{1}{r|}{154.1} & \multicolumn{1}{r|}{91.2}& \multicolumn{1}{r|}{54.0} & \multicolumn{1}{r|}{34.4} \\ \hline
		\multicolumn{1}{|r|}{100 Million} & \multicolumn{1}{r|}{\ \ \ \ \ \ \ \ 1,772.9} & \multicolumn{1}{r|}{\ \ \ \ \ \ \ \ \ 1,026.8} & \multicolumn{1}{r|}{\ \ \ \ \ \ \ \ \ \ 581.2} & \multicolumn{1}{r|}{\ \ \ \ \ \ \ \ \ 352.7}\\ \hline
	\end{tabularx}		
	\vspace{-4ex}	
\end{table}

\subsection{End-to-End Comparison with YCSB}\label{sec:ycsb}

We built four KV stores using NV-tree, FAST+FAIR,
Circ-Tree\_ls, and Circ-Tree for indexing.  
We configured the number of client threads to be 1, 2, 4, and 8, and 
 tested with two node sizes, i.e., 2KB and 4KB by loading and running
 SessionStore workload of YCSB. Figure~\ref{fig:kv-2KB} and~\ref{fig:kv-4KB} capture the 99th percentile latency with two node sizes, 
respectively.

As to the 99th percentile latencies shown in Figure~\ref{fig:kv-2KB-insert}
and Figure~\ref{fig:kv-4KB-insert} for inserting with varied client threads, the
differences among B+-tree variants are not as significant as
those with standalone B+-trees.
The reason is that, 
committing 1000B value to NVM per insertion costs much longer time than inserting a KV pair of 8B/8B
into an indexing tree and the former dominates the insertion latency.
Nevertheless, Circ-Tree still yields higher performance than NV-tree and FAST+FAIR.
With 2KB node, Circ-Tree achieved 29.3\% and 25.4\% shorter latency than
NV-tree and FAST+FAIR, respectively, with two threads.
With 4KB node, Circ-Tree achieved 23.7\% and 47.4\% shorter 99th
percentile latency than NV-tree and FAST+FAIR, respectively, with 
eight threads.

A comparison between Figure~\ref{fig:kv-2KB-insert} and 
Figure~\ref{fig:kv-4KB-insert} also indicates that, with a larger node,
1) the latency of NV-tree decreases, 2) the latency of FAST+FAIR increases, and 3) the latencies of Circ-Tree\_ls and Circ-Tree remain
consistent. Such an observation matches the observation we 
obtain with standalone B+-tree variants with larger nodes.
In the meantime, the impact of more threads is not as much as what we have seen 
with Figure~\ref{fig:multi-insert-avg}, because the long duration of committing 
1000B value per insertion overtakes the time incurred by contentions due to 
multi-threading. 

Figure~\ref{fig:kv-2KB-search} and Figure~\ref{fig:kv-2KB-update} (resp. Figure~\ref{fig:kv-4KB-search} and Figure~\ref{fig:kv-4KB-update}) capture the 
99th percentile latencies for searching and updating, respectively, with 2KB node (resp. with 4KB node). 
Search and update are both searching a B+-tree variant with additional read and write operations with NVM, respectively.
We can obtain three observations from these four diagrams.
First, in-NVM B+-tree variants yield comparable performances except NV-tree handling 
search requests issued from eight threads.
Secondly, as reading 1000B is less heavyweight than writing 100B to NVM, the latencies in Figure~\ref{fig:kv-2KB-search} (resp. Figure~\ref{fig:kv-4KB-search})
are longer than those in Figure~\ref{fig:kv-2KB-update} (resp. Figure~\ref{fig:kv-4KB-update}).
Thirdly, comparing these diagrams to Figure~\ref{fig:multi-search-avg} (i.e., 
searching standalone B+-tree variants with multi-threading workload), 
the additional costs of reading and writing data with NVM have minimized 
the performance differences among trees.

The preceding results with YCSB confirm the practical usability of Circ-Tree 
in real-world applications.

\section{Conclusion}\label{sec:conclusion}

In this paper, we
 consider a fundamental change in the design of B+-tree, 
i.e. to logically view its linear node structure in a circular fashion.  
We have built Circ-Tree with the concept of circular node 
structure and evaluated it with extensive experiments. Evaluation results 
show that Circ-Tree significantly outperforms state-of-the-art NV-tree and 
FAST+FAIR by up to 1.6$\times$ and 8.6$\times$, respectively. An end-to-end 
comparison with YCSB workload onto KV store systems based on Circ-Tree, 
NV-tree, and FAST+FAIR shows that the write latency of Circ-Tree is up 
to 29.3\% and 47.4\% shorter than that of NV-tree and FAST+FAIR, respectively.

\bibliographystyle{unsrt}
\bibliography{tree}

\begin{IEEEbiography}[{\includegraphics[width=1in,height=1.2in,clip,keepaspectratio]{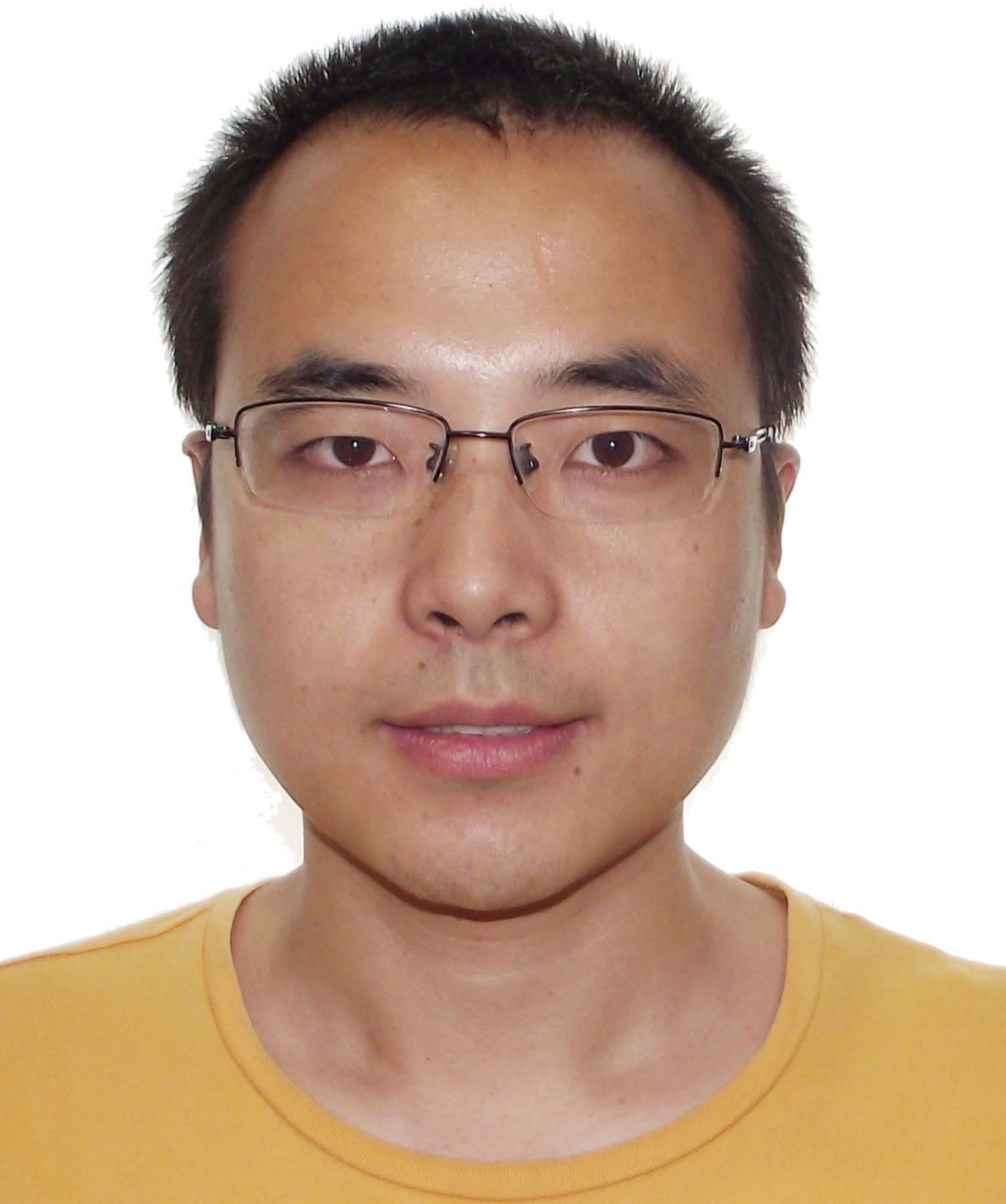}}]{Chundong Wang}
	received the Bachelor's degree in computer science from Xi'an Jiaotong University (2004-2008), and the Ph.D. degree in computer science
	from National University of Singapore (2008-2013). Currently he is
	a research fellow in Singapore University of Technology and Design (SUTD), Singapore.
	Before joining SUTD, he worked as a scientist in Data Storage Institute, A$^\star$STAR, Singapore.
	Chundong has published a number of papers in IEEE TC, ACM TOS, DAC, DATE, LCTES, USENIX FAST, etc.
	His research interests include data storage systems,
	non-volatile memory and computer architecture.
\end{IEEEbiography}

\begin{IEEEbiography}[{\includegraphics[width=1in,height=1.2in,clip,keepaspectratio]{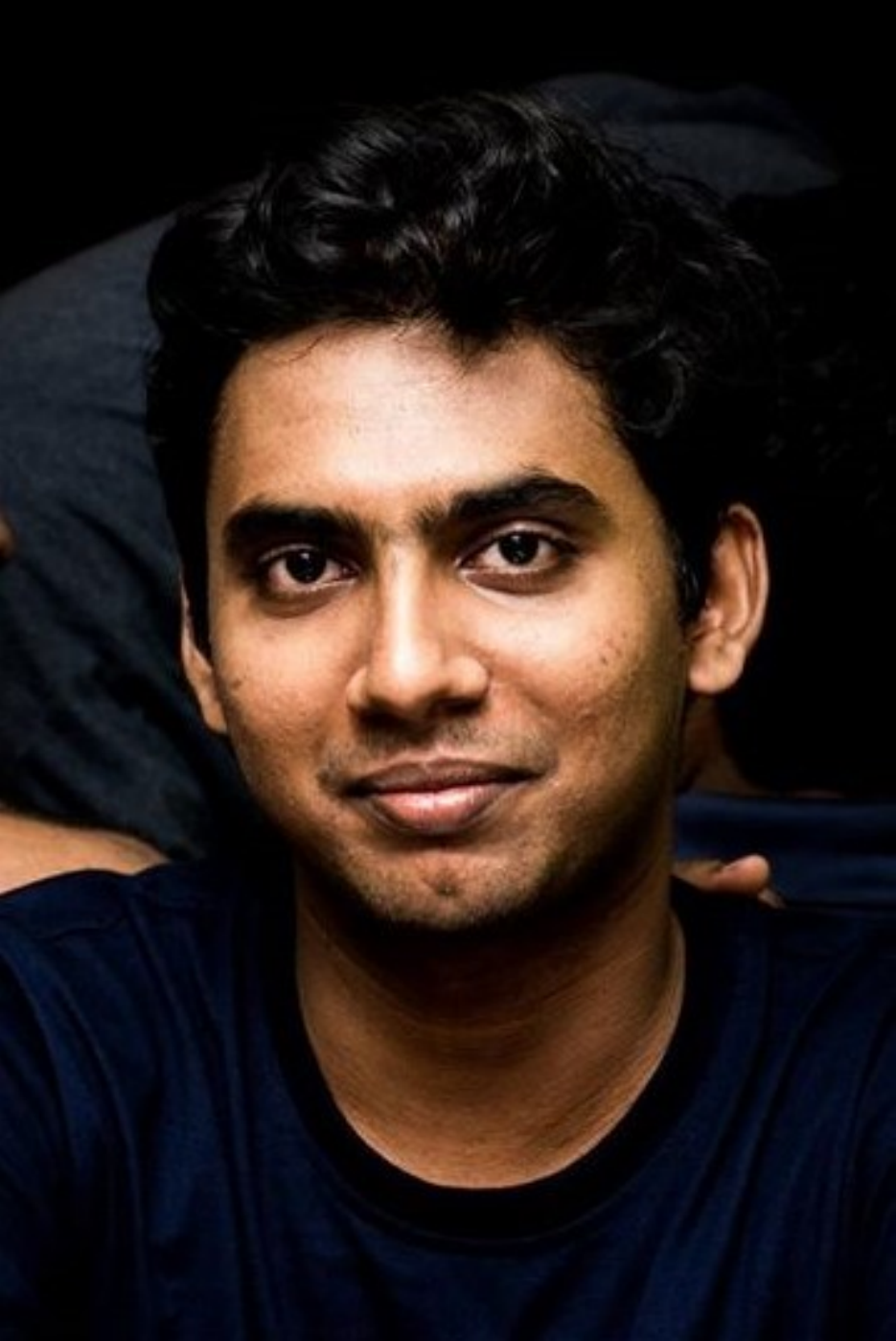}}]{Gunavaran Brihadiswaran}
is a final-year undergraduate student at the Department of Computer Science and Engineering, University of Moratuwa, Sri Lanka. 
He completed a 6-month internship in Singapore University of Technology and Design in 2018.
His research interests include bioinformatics and computational biology, parallel computing and machine learning.
\end{IEEEbiography}

\begin{IEEEbiography}[{\includegraphics[width=1in,height=1.2in,clip,keepaspectratio]{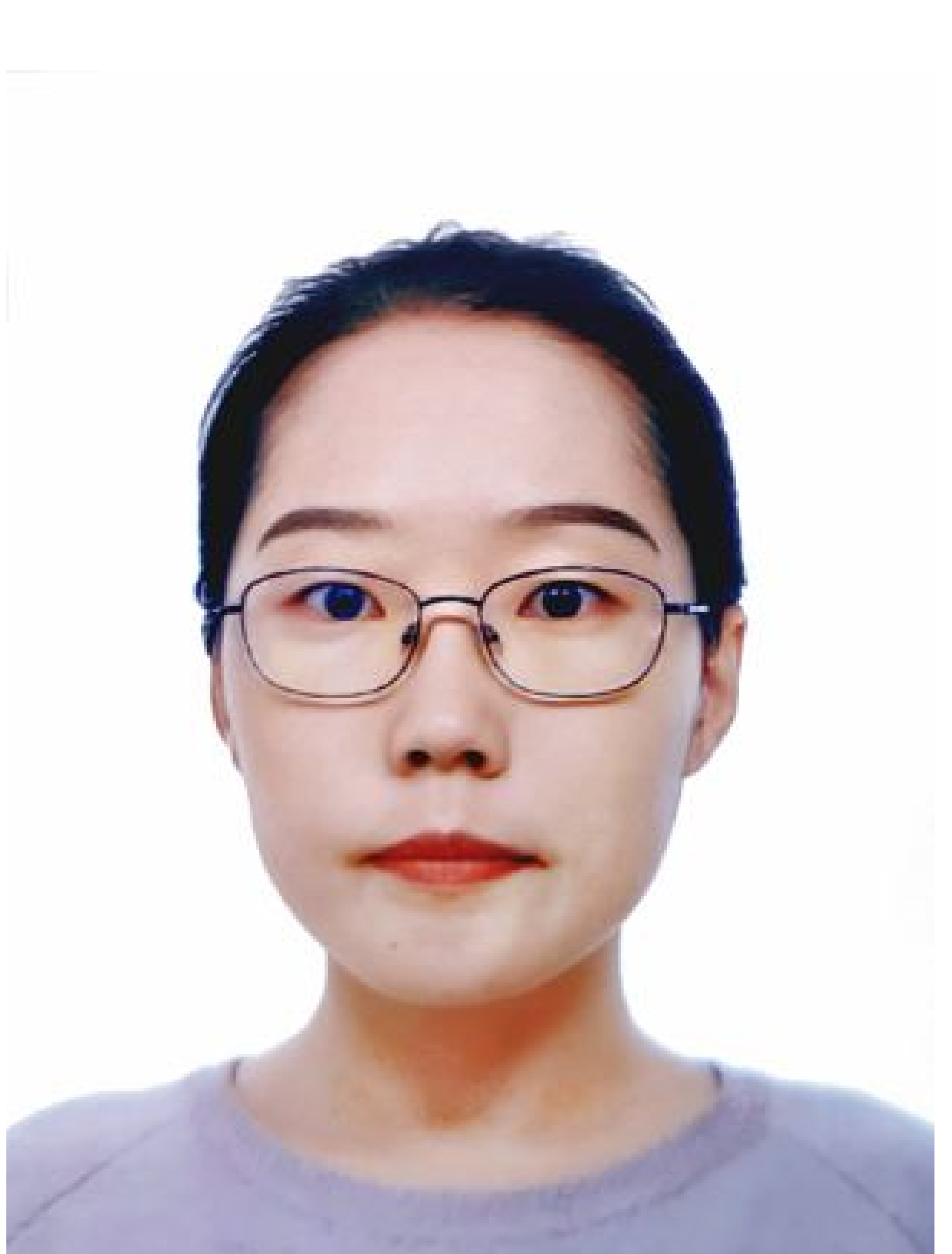}}]{Xingbin Jiang} received the B.S. degree in Electronic and Information Engineering from University of Science and Technology Beijing, China, in 2012, and the M.S. degree in Materials Engineering from the National Center for Nanoscience and Technology, Chinese Academy of Sciences, in 2015. She is currently working as a Research Assistant at the Information Systems Technology and Design (ISTD) Pillar at Singapore University of Technology and Design (SUTD).
	
	Her current research interests include IoT system security, network security, kernel security, and wireless security. 
\end{IEEEbiography}	

\begin{IEEEbiography}[{\includegraphics[width=1in,height=1.2in,clip,keepaspectratio]{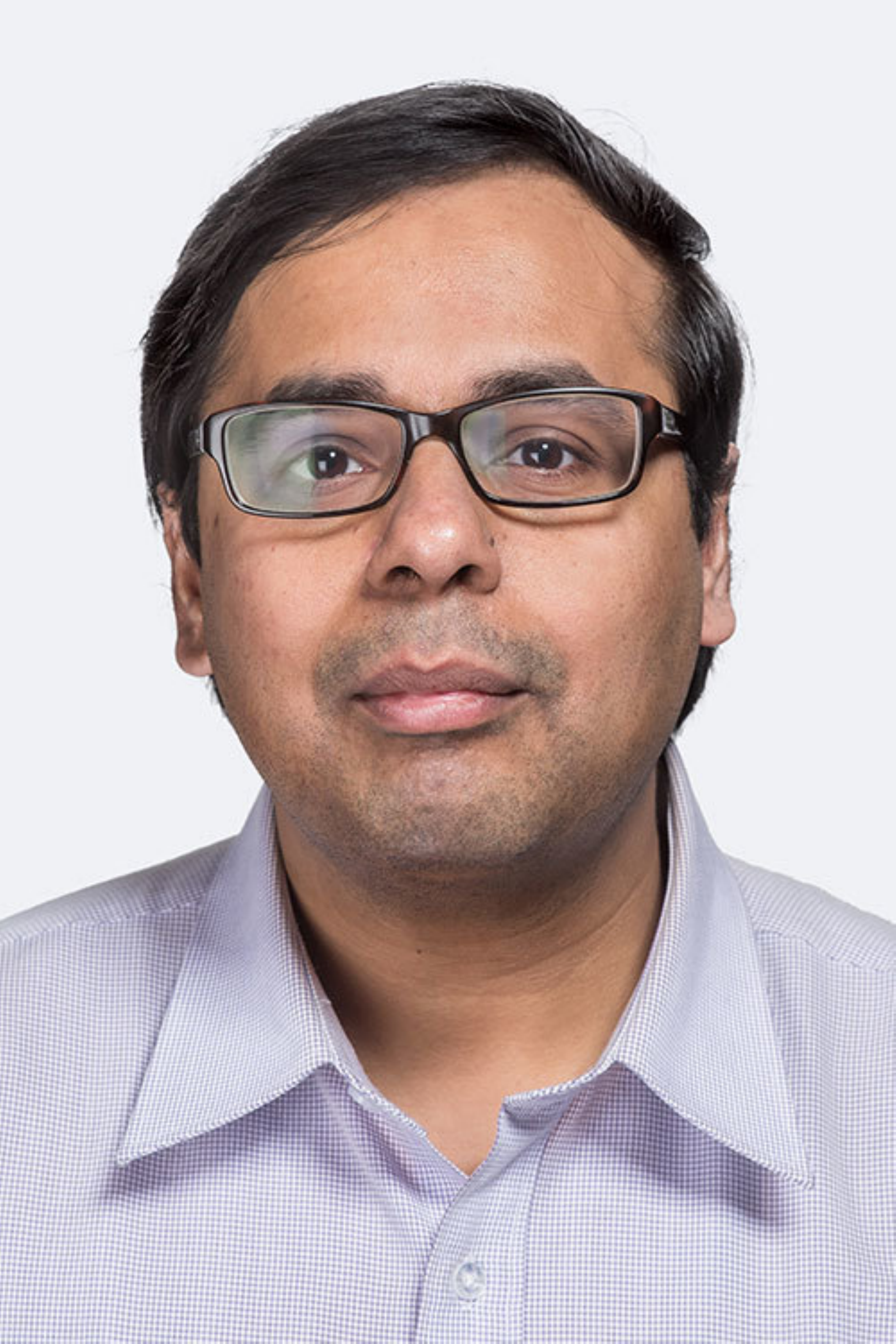}}]{Sudipta Chattopadhyay} received the Ph.D. degree in computer science from the National University of Singapore, Singapore, in 2013.
	 He is an Assistant Professor with the Information Systems Technology and Design Pillar, Singapore University of Technology and Design, Singapore. In his doctoral dissertation, he researched on Execution-Time Predictability, focusing 
	 on Multicore Platforms. He seeks to understand the influence of execution platform on critical software properties, such as performance, energy, robustness, and security. His research interests include  program analysis, embedded 
	 systems, and compilers.
	
	Mr. Chattopadhyay serves in the review board of the IEEE Transactions on Software Engineering.
\end{IEEEbiography}

\end{document}